\documentclass[12pt]{iopart}

 \expandafter\let\csname equation*\endcsname\relax 
 \expandafter\let\csname endequation*\endcsname\relax 
\usepackage{amsmath,amsfonts,amssymb}

\usepackage{graphicx,subfloat}
\usepackage{color}
\usepackage[dvipsnames]{xcolor}
\usepackage{comment}
\usepackage{appendix}
\usepackage{multirow}
\usepackage{epstopdf}
\usepackage{etoolbox}
\usepackage{hyperref}
\usepackage{grffile}
\usepackage{tikz}
\usepackage{algorithm}
\usepackage{algorithmic}

\makeatletter
\renewcommand\@appendixstar{\@@par
 \ifnumbysec 
 \@addtoreset{table}{section}
 \@addtoreset{figure}{section}\fi
 \setcounter{section}{0}
 \setcounter{subsection}{0}
 \setcounter{subsubsection}{0}
 \setcounter{equation}{0}
 \setcounter{figure}{0}
 \setcounter{table}{0}
 \def\thesection{Appendix \Alph{section}} 
 \def\theequation{\ifnumbysec
      \Alph{section}.\arabic{equation}\else
      \Alph{section}\arabic{equation}\fi}
 \def\thetable{\ifnumbysec
      \Alph{section}\arabic{table}\else
      A\arabic{table}\fi}
 \def\thefigure{\ifnumbysec
      \Alph{section}\arabic{figure}\else
      A\arabic{figure}\fi}}
\makeatother

\newcommand{\bA}{{\bf A}}

\newcommand{\ba}{{\bf a}}
\newcommand{\bb}{{\bf b}}
\newcommand{\Bf}{{\bf f}}
\newcommand{\bg}{{\bf g}}

\newcommand{\bp}{{\bf p}}

\newcommand{\Bs}{{\bf s}}

\newcommand{\bv}{{\bf v}}

\newcommand{\dsp}{\displaystyle}

\renewcommand{\d}{\text{d}}

\newcommand{\bpi}{{\boldsymbol{\pi}}}
\newcommand{\bphi}{{\boldsymbol{\phi}}}
\newcommand{\bsigma}{{\boldsymbol{\sigma}}}

\newcommand{\Dt}{h}
\newcommand{\nt}{{N_t}}

\newcommand{\mC}{\mathcal C}

\newcommand{\mE}{\mathcal E}

\newcommand{\mH}{\mathcal H}
\newcommand{\mL}{\mathcal L}
\newcommand{\mJ}{\mathcal J}

\newcommand{\mO}{\mathcal O}

\newcommand{\mR}{\mathcal R}
\newcommand{\mU}{\mathcal U}

\newcommand{\mZ}{\mathcal Z}

\newcommand{\DV}{\mathfrak D \bv}

\begin{document}
\title{Optimal response to non-equilibrium disturbances under truncated Burgers-Hopf dynamics}
\date{\today}
\author{Simon Thalabard  and Bruce Turkington} 

\address{Department of Mathematics and Statistics, University of Massachusetts, Amherst, MA 01003, USA.}
\date{\today}

\begin{abstract}
We model and compute the average response of truncated Burgers-Hopf dynamics to finite perturbations away from the Gibbs equipartition energy spectrum using a dynamical optimization framework recently conceptualized in a series of papers. Non-equilibrium averages are there approximated in terms of geodesic paths in probability space  that ``best-fit'' the Liouvillean  dynamics over a family of quasi-equilibrium trial densities.  By recasting the geodesic principle as an optimal control problem,  we solve numerically for the non-equilibrium responses using an  augmented Lagrangian, non-linear conjugate gradient descent method.
For moderate perturbations,  we find an excellent agreement between the optimal predictions and the direct numerical simulations of the truncated Burgers-Hopf dynamics. 
In this near-equilibrium regime, we argue that the optimal response theory provides an approximate yet predictive counterpart to fluctuation-dissipation  identities.

\end{abstract}
\maketitle
\newcommand{\eqspace}{\hspace{0.3cm}}

\section{Introduction}
Fluctuation-dissipation (F/D) theorems  of the first kind relate the non-equilibrium average response of systems driven away from equilibrium to corresponding two-time correlations functions computed at equilibrium \cite{kubo_statistical_2012,chandler_introduction_1987,risken_fokker-planck_1984,maes_second_2014}. 
While they constitute some of the few known exact identities of non-equilibrium statistical mechanics, it is also well known that those theorems have two major limitations~: \emph{(i)} their range of applicability is in principle restricted to infinitesimal perturbations away from equilibrium, and \emph{(ii)} they are not fully predictive~:  Two-time equilibrium  statistics  need to be  measured or computed \emph{per se} from the underlying dynamics before the desired  non-equilibrium response can be reconstructed therefrom.
Surprisingly though,  the  F/D formalism has found widespread application  in both  turbulence modeling \cite{kraichnan_irreversible_1958,kraichnan_classical_1959,biferale_fluctuation-response_2001, kraichnan_lagrangianhistory_1965, kraichnan_lagrangianhistory_1968, eyink_robert_2010} and  climate predictability \cite{leith_climate_1975,bell_climate_1980,gritsun_climate_2008,majda_challenges_2012}, two problems that  involve describing strongly out-of-equilibrium structures.

The purpose of our paper is to discuss an alternate \emph{predictive} non-equilibrium response  theory (later sometimes referred to as the `'best-fit'' theory), which both in concept and in practice  adopts a point of view opposite to the F/D framework. 
This alternate theory was introduced  in the context of deterministic dynamics \cite{turkington_best-fit_2010,turkington_optimization_2013, kleeman_nonequilibrium_2014} and qualitatively studied on a variety of prototypical problems in statistical fluid dynamics,  from the statistical homogeneization of 2D Galerkin-Euler dynamics and the truncated Burgers-Hopf dynamics \cite{kleeman_nonequilibrium_2014,turkington_coarse-graining_2016} to the single-mode energy relaxation in an inviscid GOY  shell-model \cite{thalabard_optimal_2015}. The best-fit theory approximates  single-time out-of-equilibrium averages  by selecting the closest dynamical matches  to their exact Liouvillean evolution, among a parametrized family of time-evolving trial densities.   The ``closest matches'' are  mathematically determined as the infimum of so-defined ``lack-of-fit'' actions (later defined in Section 2) and are hereafter termed  the ``optimal responses'' of the system.

In principle, the best-fit approach should be able to capture strongly non-equilibrium features. In practice, it has so far stumbled upon  the inherent difficulty of solving  explicitly the underlying non-linear optimization problem that defines the optimal response. 
 We here expose  a practical solution to this important issue~:  we describe an  ``optimal response algorithm'', that relies on an optimal control formulation of the underlying optimization problem, and  uses  an augmented Lagrangian, non-linear conjugate gradient method to optimize over the trial densities. 

 We use this algorithm to compute  the near-equilibrium optimal response of truncated Burgers-Hopf (hereafter TBH) dynamics to  finite disturbances of the energy spectrum away from equipartition. TBH  dynamics is here used as a simple prototype of a non-linear conservative deterministic dynamics with a chaotic behavior \cite{majda_remarkable_2000}. 
We note that the  subject of self-thermalization in truncated fluid models have found renewed interest over the past few years, due to its possible application to turbulence modeling \cite{cichowlas_effective_2005,krstulovic_two-fluid_2008,banerjee_transition_2014,shukla_statistical_2016}.
The statistical properties of the  TBH thermalization have been particularly scrutinized and have revealed interesting phenomenologies, from non-equipartition statistical equilibria \cite{abramov_hamiltonian_2003} to the celebrated tyger phenomenon at the onset of thermalization \cite{ray_resonance_2011,ray_thermalized_2015}.
 By contrast, we here rather focus on the late-stage properties of the statistical thermalization.  In this context,   the quasi-Gibbsian best-fit theory appears as a \emph{predictive} counterpart to  F/D type theorems,   whose range of validity is found to be comparable.

The remainder of the paper is organized as follows.
In Section 2, we  contrast the optimal and F/D description, for  the statistical response of  TBH dynamics to  a (weak) perturbation of a thermalized energy spectrum.  
 A generalized F/D theorem in the spirit of  \cite{carnevale_fluctuationresponse_1991, biferale_fluctuation-response_2001,boffetta_relaxation_2003} is swiftly  derived, and a near-equilibrium optimal response is formally defined in terms of the infimum of a well-defined lack-of-fit action.
In Section 3, we give an optimal control formulation for the optimal responses, and describe the optimal response algorithm that we use to compute them. Technicalities related to the discrete nature of the actual numerics are pushed to \ref{sec:DiscretePontryagin}. 
In Section 4, we  assess the respective predictive abilities of the optimal closure and  F/D identities with respect to direct numerical simulations (DNS) to describe the relaxation towards equipartition of  finite perturbations in the energy spectrum under TBH dynamics.
We conclude by briefly outlining interesting theoretical perspectives related to the optimal response approach.
%
%

	\section{Optimal vs Fluctuation-Dissipation responses  to initial disturbances.}			
		In this section, we contrast the conceptual framework of the best-fit theory to the F/D approach on a test-bed non-equilibrium setup : the relaxation under the TBH dynamics of an energy spectrum initially disturbed away from thermal equilibrium. We first  make precise our non-equilibrium setup, and  describe the outcomes of a  F/D-type approach. We then summarize the optimal response theory.
		\subsection{Truncated-Burgers Hopf dynamics and non-equilibrium framework}
		\paragraph{TBH and thermal equilibrium.}
			The 1D-truncated Burgers dynamics (TBH) on the  $2\pi$-torus is a relatively simple example of a chaotic non-linear conservative dynamics \cite{majda_remarkable_2000}.  It describes the non-linear evolution of a real velocity field $v(x,t) = \sum_{|l| \le K} v_l(t) e^{ilx}$ with zero spatial-mean by the projection of Burgers dynamics onto a finite set of Fourier modes, which we take to be the modes graver than a prescribed ultraviolet cutoff $K$.  Writing  $\bv = (v_l)_{1 \le l \le K}$, and using  starred symbols to denote complex conjugates,  we can write the TBH time evolution of the Fourier components $v_l(t)$ as   
		\begin{equation}
			\dot \bv(t) = \bA[\bv,\bv ^\star] ~ \text{with} ~ A_l[\bv,\bv ^\star] = 
			\begin{cases}
				\dsp
				  &\dsp -\dfrac{il}{2} \sum_{\substack{(m,n) \in [-K;K]^2\\ l+m+n=0}}v_m^\star v_n^\star  ~~~ \text{if}~|l| \le K \\
				 &0 ~\text{otherwise.}
			\end{cases}
			\label{eq:tbh}
		\end{equation}		
The single-time statistics of $\bv$ are then fully determined from the evolution of the densities $p(\bv,t)$ under the Liouville operator $\mL$ as :
			\begin{equation}
				\partial_t  p + \mL\; p  =0 ~ \text{with} ~ \mL = \sum_{l=1}^K \left( A_l\partial_{v_l} + A_l^\star \partial_{v_l^\star}\right) =  \bA \cdot \nabla_\bv + \bA^\star \cdot \nabla_\bv^\star, 
			\label{eq:liouville}	
			\end{equation}
where the detailed Liouville property $\partial_{v_l} A_l = 0$ is used to deduce (\ref{eq:liouville}) from (\ref{eq:tbh}).  The last equality in the previous equation is used to define some convenient  shorthand notations~: Partial derivatives with respect to the $v_l$'s and $v_l^\star$'s are complex derivatives, the nabla notation  means  $\nabla_\bv = (\partial_{v_l})_{1\le l \le K} $ and the scalar product is defined as $\Bf \cdot \bg = \sum_{l=1}^K f_l g_l$.

		The single-time statistics of the equilibrium distributions are obtained as functionals of the dynamical invariants of  (\ref{eq:tbh}), primary among which is the kinetic energy $\mE[\bv] = \sum_{1\le l\le K} v_l v_l^\star$.  The latter yields the Gibbs equipartition  distribution, defined in terms of the inverse temperature $\beta$ , which we write as  $p_\beta =  \mZ_\beta^{-1}\exp \left(-\beta \mE\right)$  with $\mZ_\beta= (\beta/\pi)^K$, the canonical partition function. The inverse temperature $\beta$  determines the average energy contained at scale $l$,  namely $ \langle v_l^2\rangle = \beta^{-1}$.
		
		\paragraph{Non-equilibrium setup.}
		In this work, we consider the following non-equilibrium protocol. At time $t=0$, we draw an ensemble of statistically homogeneous fields $\bv$, sampled from a quasi-Gibbsian  distribution, which may be thought of as a ``disturbed equipartition state''. The latter is  defined in terms of a non-uniform inverse temperature vector $\bb = (b_l)_{1\le l \le K}$, namely $ p( \cdot,t=0)= p_\bb(\cdot)$ with
		\begin{equation}
			p_\bb[\bv] =  \prod_{l=1}^K p_{b_l}(v_l), ~~\text{and}~~ p_{b_l} (v_l) = \dfrac{b_l}{\pi} \exp \left(-b_l v_l v_l^\star\right).
		\label{eq:quasigibbs}
		\end{equation}
		Because of the chaotic nature of the dynamics, it is reasonable to expect that at long time the quasi-Gibbsian distribution relaxes towards the equipartition distribution, with inverse temperature $\beta = K/\sum_{l=1}^K b_l^{-1}$ \cite{abramov_hamiltonian_2003}.   As a reminder,  the time-dependent non-equilibrium averages of any observable $\mO $ are defined as  
		\begin{equation}
		\begin{split}
			& \langle \mO \rangle_t = \int \DV \;  \mO (\bv) p(\bv,t), ~\text{~where~} p(\cdot,t) = e^{-t\mL} p_\bb(\cdot) , \\
		 & \text{~and~} \int \DV = \prod_{l=1}^K \int_{\mR^2} \d \Im v_l \; \d \Re v_l .
		\label{eq:noneqaverage}
		\end{split}
		\end{equation}
 The non-equilibrium averages can also be formally written in terms of forward propagators as 
		\begin{equation}
			\begin{split}
				&\langle \mO \rangle_t = \int \DV_0 \;  \mO (t| \bv_0) p_\bb(\bv_0) 
				,\text{~where}~\mO (t| \bv_0) = \int \DV \;\mO[\bv] P(\bv,t|\bv_0) p_\bb(\bv_0)\\
				 &\text{~and ~}  P(\bv,t|\bv_0) = e^{-t\mL} \delta(\bv-\bv_0).
			\end{split}
			\label{eq:noneqaverage2}
		\end{equation}
Equilibrium averages, which we later simply denote as  $\langle \cdot \rangle_\beta$, are obtained by taking all $b_l  = \beta$ in (\ref{eq:noneqaverage}), and using the invariant measure property : $\int \DV_0 P(\bv,t|\bv_0) p_\beta(\bv_0) = e^{-t\mL} p_\beta(\bv) = p_\beta(\bv)$.
		Both the F/D and the best-fit approaches aim at describing the evolution of the non-equilibrium averages $\langle \cdot \rangle_t$.
	
		\subsection{Generalized Fluctuation-Dissipation identity.}
		For our specific set-up, a linear F/D estimate is derived as in \cite{carnevale_fluctuationresponse_1991,biferale_fluctuation-response_2001,boffetta_relaxation_2003}.  It expresses  the deviation $\Delta \langle \mO \rangle = \langle \mO \rangle_t - \langle \mO \rangle_\beta$ from equilibrium in terms of the response functions $R_\mO^l$ and the initial perurbation in the energy spectrum $1/b_l -1/\beta$ as 
		\begin{equation}
			\begin{split}
			& \Delta \langle \mO \rangle \simeq  \sum_{l=1}^K R_\mO^l(t) (b_l^{-1}-\beta^{-1} )
			\text{~with~}  R_\mO^l(t) =-\beta^2 \left \langle  \mO(t|\bv_0) \left. \dfrac{\partial \log p_{\bb}[\bv_0]}{\partial b_l}\right|_{b_l=\beta} \right \rangle_\beta,\\
			& \text{and explicitly } R_\mO^l(t) =- \beta \left \langle  \mO(t|\bv_0) \left(1- \beta |v_{0,l}|^2  \right) \right \rangle_\beta.
			\end{split}
			\label{eq:FD}
		\end{equation}		
		In particular, the F/D estimate for the  energy contained at wavenumber  $k$ is obtained by setting $ \mO(\bv)=  |v_k|^2$ in  (\ref{eq:FD}), and reads
		\begin{equation}
			\begin{split}
			& \Delta \langle| v_k|^2 \rangle \simeq  \sum_{l=1}^K R_k^l(t) (b_l^{-1}-\beta^{-1} )
			\text{~with~}  R_k^l(t) =\beta^2 \left \langle | v_k|^2(t)| v_{l}|^2(0) \right \rangle_\beta -1.
			\end{split}
			\label{eq:FDestimateEnergy}
		\end{equation}		
		The derivation of (\ref{eq:FD}) is straightforward, and obtained by expanding to first order in $|\bb^{-1} -\beta^{-1} | \ll 1 $ the following identity, that stems from (\ref{eq:noneqaverage})~:
		\begin{equation}
			\Delta \langle \mO\rangle = \left \langle \mO(t | \bv_0) \;F(\bv_0,\bb , \beta) \right \rangle_\beta ~~\text{with} ~~F(\bv_0,\bb,\beta) = \dfrac{p_\bb[\bv_0] - p_\beta[\bv_0]}{p_\beta[\bv_0]}.
		\end{equation}
		Clearly, the central objects of the F/D approach are the response functions, \emph{i.e.} two-time equilibrium correlations~:  the \emph{r.h.s} of  (\ref{eq:FD}) provides a formula to reconstruct the non-equilibrium averages  from the latter, provided the deviations from equilibrium are small enough. 
		
		\subsection{A brief exposition of the optimal response theory.}
\begin{figure}
			\centering
			\includegraphics[width=0.69\textwidth]{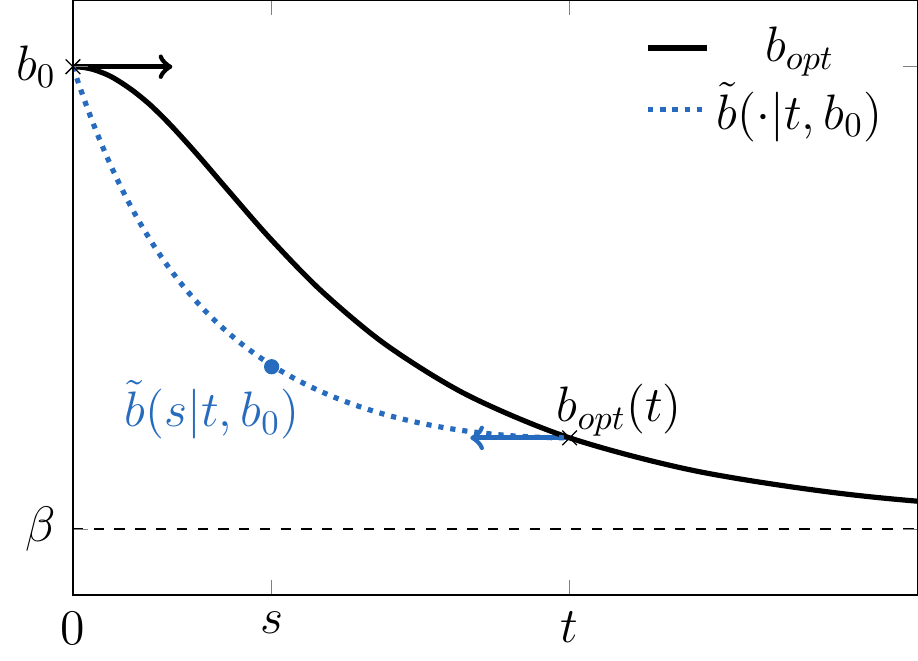}
			\caption{A one-dimensional  illustration of the quasi-Gibbsian optimal response $\bb_{opt}(t)$, defined as the time enveloppe of the shadow evolutions $\tilde  \bb(.|t)$ (see text for definitions).}
			\label{fig:FWDcartoon}
		\end{figure}

\paragraph{Concept.}
On the other hand, the cornerstone  of the best-fit theory is the Liouvillean evolution itself (\ref{eq:liouville}).
The philosophy is to model  the non-equilibrium averages in terms of explicitly computable ``trial averages'', whose dynamics in probability shadow the  true Liouvillean evolution  (\ref{eq:liouville}).
In our case, the simplest prescription is to assume that the non-equilibrium densities remain  quasi-Gibbsian (\ref{eq:quasigibbs}) throughout the relaxation. More mathematically, this means the following \emph{Ansatz}: there exists an optimal quasi-Gibbsian path in probability space, namely a smooth dynamical path $[\bb^{opt}] =  \left\lbrace \bb^{opt}(t) \right \rbrace_{t=0}^\infty$, and associated quasi-Gibbsian density evolution $\rho_{\bb^{opt}}$ such that  
		\begin{equation}
			\langle\cdot  \rangle_t \simeq \langle \cdot \rangle_{\bb^{opt}(t)} = \int \DV \rho_{\bb^{opt}(t)} (\bv)\cdot
			\label{eq:approx}
		\end{equation}
The best-fit theory provides a systematic framework to determine $[\bb^{opt}]$  as a best-fit to the actual  Liouvillean dynamics among all the quasi-Gibbsian paths.
Different versions of the theory exist. We here use the one previously described in \cite{turkington_best-fit_2010}, that we may describe as a forward, non-stationnary best-fit theory. 
%
\paragraph{Lack-of-fit cost.}
For each feasible path $[\bb]$, we define the following time-dependent \emph{Liouville residual }:
\begin{equation}
	\mR_{[\bb],t} = \left( \partial_t + \mL \right) \log \rho_{\bb(t)},
	\label{eq:LiouvilleResidual}
\end{equation}
the average square of which is interpreted as a  \emph{lack-of-fit Lagrangian} $L_{lof}$ \cite{turkington_optimization_2013}.
The discrepancy up to time $t$ between the true p.d.f and the quasi-Gibbsian evolution is then measured in terms of a \emph{lack-of-fit cost function} $J_{lof}$  defined as ~:
\begin{equation}
	J_{lof}([\bb],t) = \int_0^t \d s\;L_{lof}\left[\bb(s), \dot \bb(s) \right] \;\; ~\text{with} ~L_{lof}  \;\; =\dfrac{1}{2} \left\langle \mR^2_{[\bb],t}\right \rangle_{\bb(t)}.
\end{equation}
It will also prove useful to work with the Legendre transform of the lack-of-fit Lagrangian, a quantity that we naturally call the \emph{lack-of-fit Hamiltonian} : 
	\begin{equation}
		\mH_{lof}(\bb,\bpi) = \dot \bb \cdot \bpi - L_{lof} \text{~with~} \bpi = \nabla_{\dot \bb} L_{lof}.
	\label{eq:lofHamilton}
	\end{equation}
Choosing the inverse temperature vector $\bb$ as the state variable, the latter represents the inverse of the energy spectrum : $\langle v_l^2 \rangle_{\bb}  = 1/b_l$. The conjugate variable  $\bpi$ has in that case the dimensions of an energy dissipation, and $\bpi_l$  represents the energy transfer function  at wavenumber $l$.
%
%
%
\paragraph{Principle  of least dynamical discrepancy.}
We now  advocate the use of what we may  call a  ``principle of least dynamical discrepancy'', to define the optimal cost  as the cost that minimizes the Liouvillean discrepancy among all the quasi-Gibbsian paths, namely~: 
\begin{equation}
	J_{lof}^{opt}(\bb_0,t) = \underset{\substack{[\bb] :\bb(t=0)=\bb_0}}{\inf} J_{lof}([\bb],t).  
\label{eq:optimalcost}
\end{equation}
It is formally determined as the solution to the ``backward'' Hamilton-Jacobi equation : 
\begin{equation}
	\begin{split}
		&\partial_t J + \mH_{lof} \left( \bb_0,- \nabla_{\bb_0} J \right) = 0 \text{~with initial condition~} J(\bb_0,t=0)=0.\\	
	\end{split}
	\label{eq:nonstatHJ}
\end{equation}
For a fixed time $t$,  the free-end  optimization  problem (\ref{eq:optimalcost}) is solved by  the path  $[\tilde \bb(.|t)]$, which we hereafter call a ``shadow optimal path''.  Its time-evolution up to  time $t$ is determined by the Hamilton equations associated to the lack-of-fit Hamiltonian (\ref{eq:lofHamilton}), and satisfies a two-end boundary conditions :  $\tilde \bb(0|t) = \bb_0$ and $ \tilde \bpi(t|t) = 0$ . While each shadow optimal response represents the ``best-fit  up to time $t$'' to the actual Liouvillean dynamics (\ref{eq:liouville}),  there is \emph{a priori} no good reason to single out a specific time $t$ to define the optimal response of the system. We therefore define the latter as  the time enveloppe of the shadow paths~:
\begin{equation}
\bb_{opt}(t) = \tilde \bb (t|t,\bb_0)  ~\text{with}~ \tilde \bb (.|t,\bb_0) =  \underset{[\bb]:\bb(t=0)=\bb_0}{\arg\min}\; J_{lof}[[b],t].
\label{eq:shadows}
\end{equation}
Equations (\ref{eq:nonstatHJ}) and (\ref{eq:shadows}) then entirely prescribe the optimal response. The definitions are illustrated on Figure \ref{fig:FWDcartoon}.
\paragraph{Comment~: Stationary vs non-stationary response.}  In previous papers, a special attention was given to  a ``stationary'' optimal response,  defined as $ \bb_{opt}^{stat}(t) =\tilde \bb(t|\infty)  $. For large times, the stationary response coincides with the optimal response (\ref{eq:shadows}). For short time, however, the stationary response poorly models  the true dynamics. For example, in our particular case, where the initial ensembles are taken to be quasi-Gibbsian,  the average initial (true) dissipation of the energy  is vanishing.
On the other hand,  the initial energy dissipation of the stationary optimal path is determined by the formula $\pi_0^{stat} = -\nabla_{\bb_0} J(\bb_0,\infty)$. For the quasi-Gibbsian Ansatz, this quantity is  non-zero unless the initial perturbation is already at equilibrium. 
By contrast, the optimal response (\ref{eq:shadows}) defined in terms of the shadow enveloppe is ``non-stationary'', and allows the model to accommodate both the desired initial condition and the requirement of  reaching equilibrium when $t\to \infty$.
\paragraph{Lack-of-fit Hamiltonian for TBH.}
To complete the specification of the optimal path, it only remains to compute the lack-of-fit Lagrangian and associated lack-of-fit Hamiltonian relevant for the TBH dynamics.
The calculation is straightforward. It consists in plugging the Ansatz (\ref{eq:quasigibbs}) into the definition of the  Residual (\ref{eq:LiouvilleResidual}), and tediously compute its averaged square with respect to the quasi-Gibbsian density. Similar calculations being  detailed in \cite{thalabard_optimal_2015,turkington_coarse-graining_2016}, we here only give the final results. The lack-of-fit Lagrangian reads :
\begin{equation}
	L_{lof}(\bb,\dot \bb) = \sum_{l=1}^K \dfrac{\dot b_l^2}{2 b_l^2} + \mU[\bb] \text{~~with~~} \mU[\bb] = 
	\sum_{l=1}^K \sum_{\substack{(m,n) \in [-K;K]^2\\ l+m+n=0}}\left( \dfrac{l^2 b_l}{2 b_mb_n} + \dfrac{mn}{b_l} \right), 
\label{eq:TBHLofLagrangian}
\end{equation}
from which we obtain the  lack-of-fit Hamiltonian as 
\begin{equation}
	\mH_{lof}(\bb,\bpi) = \sum_{l=1}^K \dfrac{\pi_l^2 b_l^2}{2} - \mU[\bb]. %
	\label{eq:TBHLofHamiltonian}
\end{equation}
We use the convention $b_{-l} = b_l$ in (\ref{eq:TBHLofLagrangian}).
By analogy with classical mechanics, we refer to  $\mU[\bb]$ as the the lack-of-fit potential.
It can be checked by a direct calculation that it indeed vanishes at equilibrium, $\mU[\bb= \beta] =0$.
Slightly anticipating Section \ref{sec:Pontryagin}, we can already observe that $\beta$ need not be specifically tied to the energy of the initial ensemble. For \emph{any} constant vector $\bb = b$, we in fact have $\mU[\bb= b] = \dfrac{\beta}{b} \mU[\bb = \beta] =0$. 

\paragraph{Estimates of non-equilibrium averages.}
	In principle, the approximation (\ref{eq:approx}) makes the best-fit theory predictive. That is,  the deviations from equilibrium are now estimated in terms of single-time averages : 
		\begin{equation} 
			\Delta \langle \mO \rangle_t = \left\langle \mO(\bv) \;F(\bv,\bb_{opt}(t) , \beta) \right\rangle_\beta ~~\text{with} ~~F(\bv,\bb_{opt},\beta) = \dfrac{p_{\bb_{opt}(t)}[\bv] - p_\beta[\bv]}{p_\beta[\bv]},
			\label{eq:optimalestimateObservable}
		\end{equation}
	where $\bb_{opt}$ is the optimal response of the TBH dynamics, as determined from the shadow paths  (\ref{eq:shadows}) and the Hamilton-Jacobi evolution (\ref{eq:nonstatHJ}) for the quasi-Gibbsian lack-of-fit Hamiltonian (\ref{eq:TBHLofHamiltonian}).
	In particular, the evolution of the energy spectrum is then simply estimated by	
		\begin{equation} 
			\Delta \langle |v_k|^2 \rangle_t = \dfrac{1}{b_{opt,k}(t)} - \dfrac{1}{\beta}.
			\label{eq:optimalestimateEnergy}
		\end{equation}	
	This single-time formula needs to be constrasted to the F/D estimate (\ref{eq:FDestimateEnergy}), which involves two-time quantities.
	
	  In practice, however, solving the  Hamilton-Jacobi equations  is an inherently difficult task, and is in general beyond the reach of  analytical means. The difficulty  comes from the nonlinear nature of the underlying optimization problem. In the past, further approximations have been advocated, such as  perturbation expansions and  mean-field approximations, in order to provide a closed  set of ordinary differential equations for the evolution of the optimal paths \cite{kleeman_nonequilibrium_2014,turkington_coarse-graining_2016}.  Such solutions are not entirely satisfactory to assess the predictive skills of the optimal theory \emph{per se}, as it is then not clear how to disentangle the discrepancy due to the quasi-Gibbsian Ansatz from the discrepancy due to our inability to provide a clear cut solution to the Hamilton-Jacobi equation. This limitation can however  be overcome by numerics. Optimization algorithms can indeed be implemented to compute the shadow responses directly from (\ref{eq:optimalcost}), hereby providing a way to determine the optimal responses. Their description is the subject of the next section.

\section{The optimal response algorithm.}
\label{sec:Pontryagin}
In this section we implement an iterative method to compute the optimal response numerically. The shadow paths (\ref{eq:shadows}) are determined directly by minimizing the optimal cost, rather than  solving   the Hamilton-Jacobi equation explicitly.
To achieve such a task, we first reformulate the optimization problem (\ref{eq:nonstatHJ}) in terms of an optimal control problem. We then outline a non-linear descent algorithm and argue that we need to resort to an augmented Lagrangian approach, in order  to  enforce energy conservation along the optimal relaxation. 
	\subsection{Optimal control formulation of the shadow paths. }
	 Minimizing the cost (\ref{eq:optimalcost}) over the trial paths $ [\bb]$ is equivalent to optimizing over the controls $[\bsigma]$ the following objective functional : 
		\begin{equation*}
			\mJ[[\bsigma],t] = \int_0^t\d s \;  L_{lof}\left(\bb(s),\bsigma(s)\right) \text{~subject to ~} %
			\begin{cases}
				& \bb[t=0] = \bb_0	\\
				&  \dot \bb = \bphi(\bb,\bsigma) \text{~with ~}  \phi_l(\bb,\bsigma) = b_l \sigma_l	\\
			\end{cases} 
		\end{equation*} 
We use the notation $[\bsigma]$ to emphasize that the control is a function $[0,t] \to \mathbb R^K$. 
The optimal control formulation enslaves the path $[\bb]$  to the control, in the same way that the Lagrangian formulation (\ref{eq:optimalcost}) ties its time derivative to the path $\bb$.
Necessary conditions for optimality can then  be obtained with  the method of Lagrange multipliers~: we introduce the co-state $[\bpi]$ to enforce the dynamical constraint and  we now look for the extremal points of the following extended objective function :
		\begin{equation}
			\begin{split}
			&\tilde \mJ[[\bsigma],t] = \int_0^t\d s \;\left(  \dot \bb(s) \cdot \bpi (s) - \mH_P(\bb(s),\bpi(s),
			\bsigma(s))\right),\\
		& \text{where ~} \mH_P = \bphi \cdot \bpi - L_{lof}(\bb,\bsigma)  ~\text{defines the Hamilton-Pontryagin function}.
			\end{split}
			\label{eq:extentedlag}
		\end{equation}
		
Following the terminology found in the optimization literature \cite{bryson_applied_1979, nocedal_numerical_2006},  we hereafter denote the arguments of the Hamilton-Pontryagin function as the state ($\bb$) the co-state ($\bpi$) and the control ($\bsigma$).
Let us now fix the time $t$ and the initial state $\bb_0$.  The variations of the objective function induced by  infinitesimal admissible variations of its functional arguments read :  
\begin{equation}
	\begin{split}
		\delta \tilde \mJ = \int_0^t \d s \; & \left\lbrace  \left(\dot \bb(s) - \nabla_\bpi \mH_P \right) \cdot \delta \bpi(s)  - \left(  \dot\bpi(s) + \nabla_\bb \mH_P \right) \cdot \delta \bb(s) \right.\\%
		& \left.- \nabla_\bsigma \mH_P  \cdot \delta \bsigma(s) \right\rbrace 
		+ \delta \bb(t) \cdot \bpi(t), 
	\end{split}
	\label{eq:variation}
\end{equation}
and  the extremal paths of the extended action therefore solve the following optimality conditions~:
	\begin{equation}
		\begin{split}
			\dot\bb = \nabla_{\bpi} \mH_P = \bphi(\bb,\bsigma)  ~~~~\text{~with~} \bb(0) = \bb_0   \hspace{2cm}  &\text{(state equation)},\\
			 \dot\bpi =- \nabla_{\bb} \mH_P  ~~~~ \text{~with~} \bpi(t) = 0 \hspace{2cm}  &\text{(co-state equation)},\\
	\text{and~~~~~}		 \nabla_\bsigma \mH_P  = 0  \hspace{2cm} &\text{(optimal control)}.
		 \end{split}
	\label{eq:optimal}
	\end{equation}
In other words, the shadow paths are obtained by solving two ordinary differential equations (the state and the costate equations), provided that the optimal control $[\bsigma]$ is known.

Let us observe that in the optimal control formulation, the state variable is still the inverse temperature vector, so that the co-state still represents the rates of energy transfer. The control $\bsigma$ has dimension of a neg-entropy production vector. That is, $\sigma_k$(t) represents minus the entropy production of the single-mode marginal at wave-number $k$ of the trial density.
%
%
\subsection{Iterative method to solve for the optimal control.}
Equations (\ref{eq:variation}) and (\ref{eq:optimal}) suggest an iterative descent method to solve numerically for the optimal  control : namely define a sequence of estimates for the control $[\bsigma^{(k)}]$ (and associated state and co-state), that converges to an optimal control $[\bsigma_{opt}]$ when $k \to \infty$. A popular choice is to use one of the many non-linear conjugate gradient type algorithms, whose philosophy is the following.

\begin{itemize}
	\item We start from  an initial guess for the control. It can  for example be $[\bsigma^{(0)}] = 0$, or a previously computed optimal control  up to some time $t^\prime < t$.
	\item Given an estimate  $[\bsigma^{(k)}]$ for the control,  we compute the estimates  $[\bb^{(k)}]$ for the state and   $[\bpi^{(k)}]$ for the co-state  by the forward integration of the state equation, and the backward integration of the co-state equation, respectively~ :
	\begin{equation}
		\begin{split}
			& \dot\bb^{(k)}(s) =\bphi\left( \bb^{(k)},\bsigma^{(k)}\right)  ~~~~~~~~~~~~~\text{~from ~} \bb^{(k)}(0) = \bb_0 ,\\
			  \text{and}~~~&\dot\bpi^{(k)}(s) =- \nabla_{\bb^{(k)}(s)} \mH_P\left[\bb^{(k)},\bpi^{(k)},\bsigma^{(k)}\right]  ~~~~~~\text{~from~} \bpi^{(k)}(t) = 0.
		 \end{split}
	\label{eq:optimalestimates}
	\end{equation}

	\item  Various schemes provide the  update  $[\bsigma^{(k+1)}]$ from $[ \bsigma^{(k)}]$.
		To first order, the corresponding variation  of the  objective cost would then read : 
	\begin{equation}
		\begin{split}
		\delta \tilde \mJ = - \int_0^t \d s \; \nabla_{\bsigma^{(k)}}  \mH_P\left[\bb^{(k)},\bpi^{(k)},\bsigma^{(k)}\right]  \cdot \left( \bsigma^{(k+1)}(s) -  \bsigma^{(k)}(s) \right).
		\end{split}
	\label{eq:itervariation}
	\end{equation}
	 The non-linear conjugate gradient method consists in taking $ [\bsigma^{(k+1)}]$ as a carefully chosen linear combination of the functional gradient formally defined through (\ref{eq:itervariation}) with an iteratively defined   search direction $[\bp^{(k)}]$.  Here,  we use  the so-called `` Polak-Ribi\`ere$^+$'' formula to update the search directions, and take $[\bsigma^{(k+1)}] =[\bsigma^{(k)}]  + \alpha^{(k)} [\bp^{(k)}]$, where  $\alpha^{(k)}$ is determined via a line-search algorithm that  ensures the so-called Wolfe conditions to be satisfied (see \cite[chapter 5]{nocedal_numerical_2006} and the details in \ref{sec:DiscretePontryagin}).
\end{itemize}
In practice, the previous algorithm is guaranteed to converge towards the desired optimal control provided that the objective cost function can be computed exactly, along with its  functional gradient.  Standard Runge-Kutta  algorithms can in principle be expected to give a good approximation of the successive state and co-state estimates, thereby obtaining reasonable approximations of the gradients.
In order to reduce the numerical flaws, however,  we find it safer to approximate the objective cost function by a discrete time counterpart, and to  perform an \emph{exact} descent (up to machine precision). The discrete formulation being more technical than enlightening, we refer the reader interested in the implementation details to \ref{sec:DiscretePontryagin}.

	\subsection{Mean-field behavior and statistical energy conservation.}
\begin{figure}
		\begin{minipage}{0.59\textwidth}
			\centering
			\includegraphics[width=\textwidth]{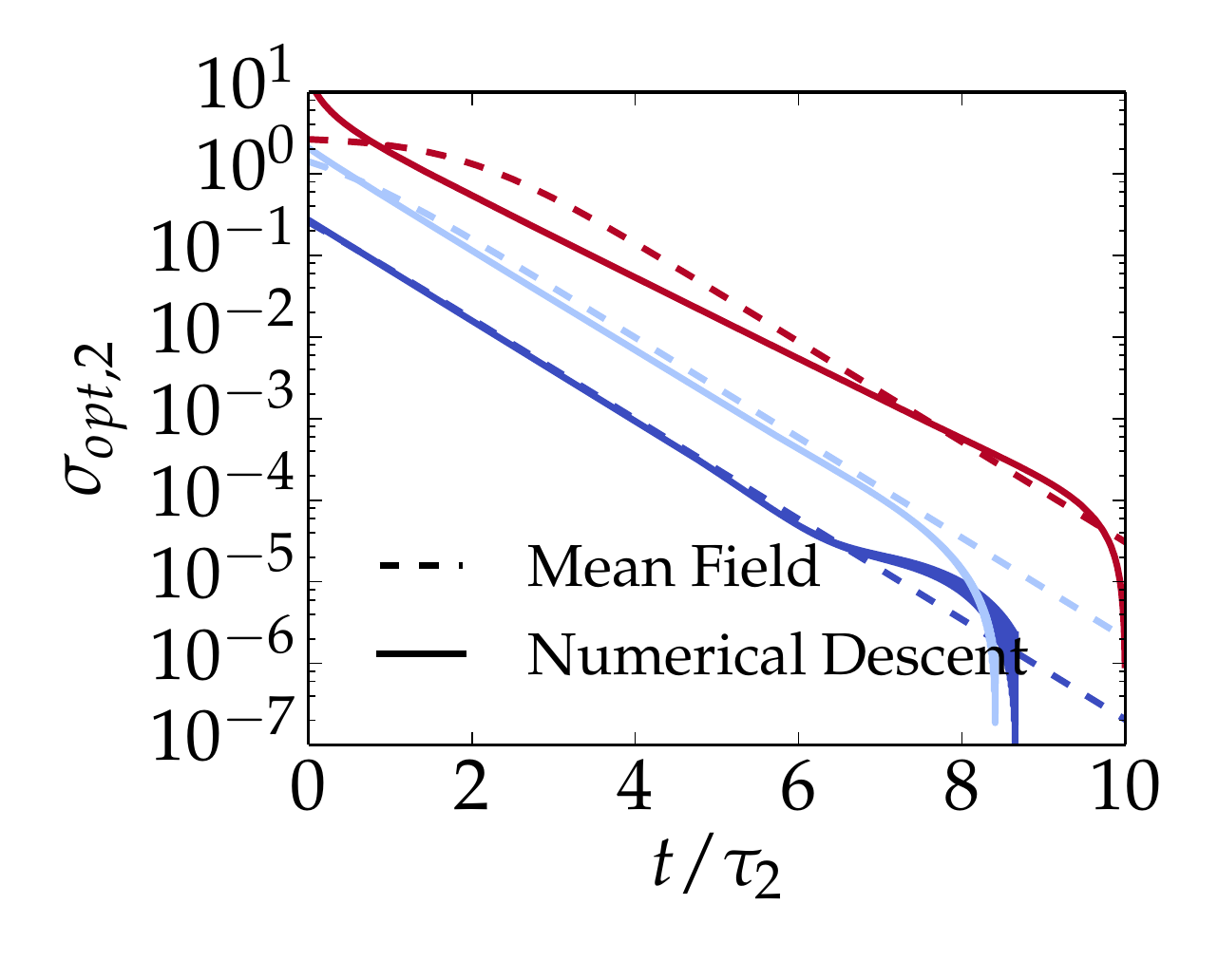}\\
		\end{minipage}
		\begin{minipage}{0.4\textwidth}
			\centering
			\includegraphics[width=\textwidth]{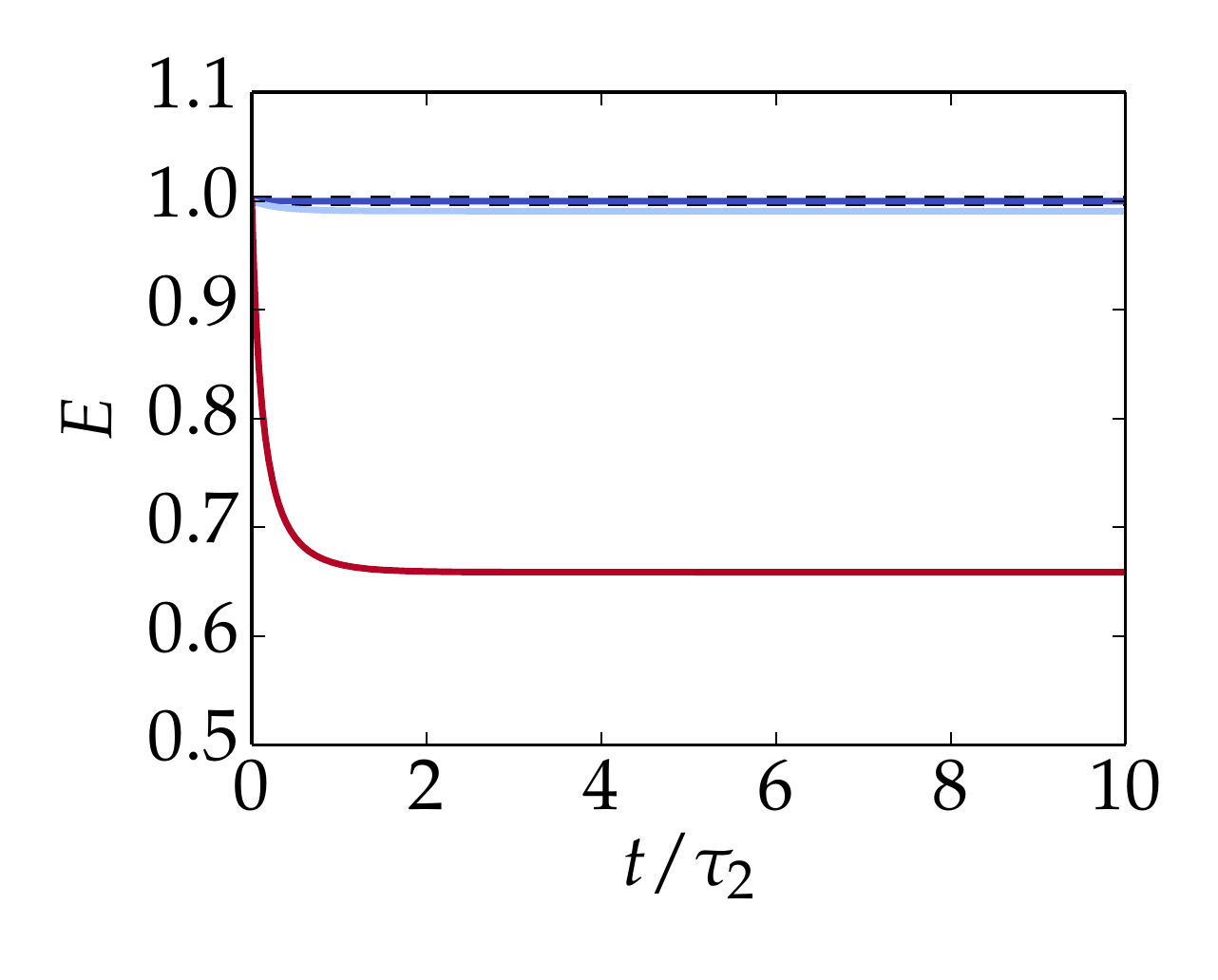}
			\includegraphics[width=\textwidth]{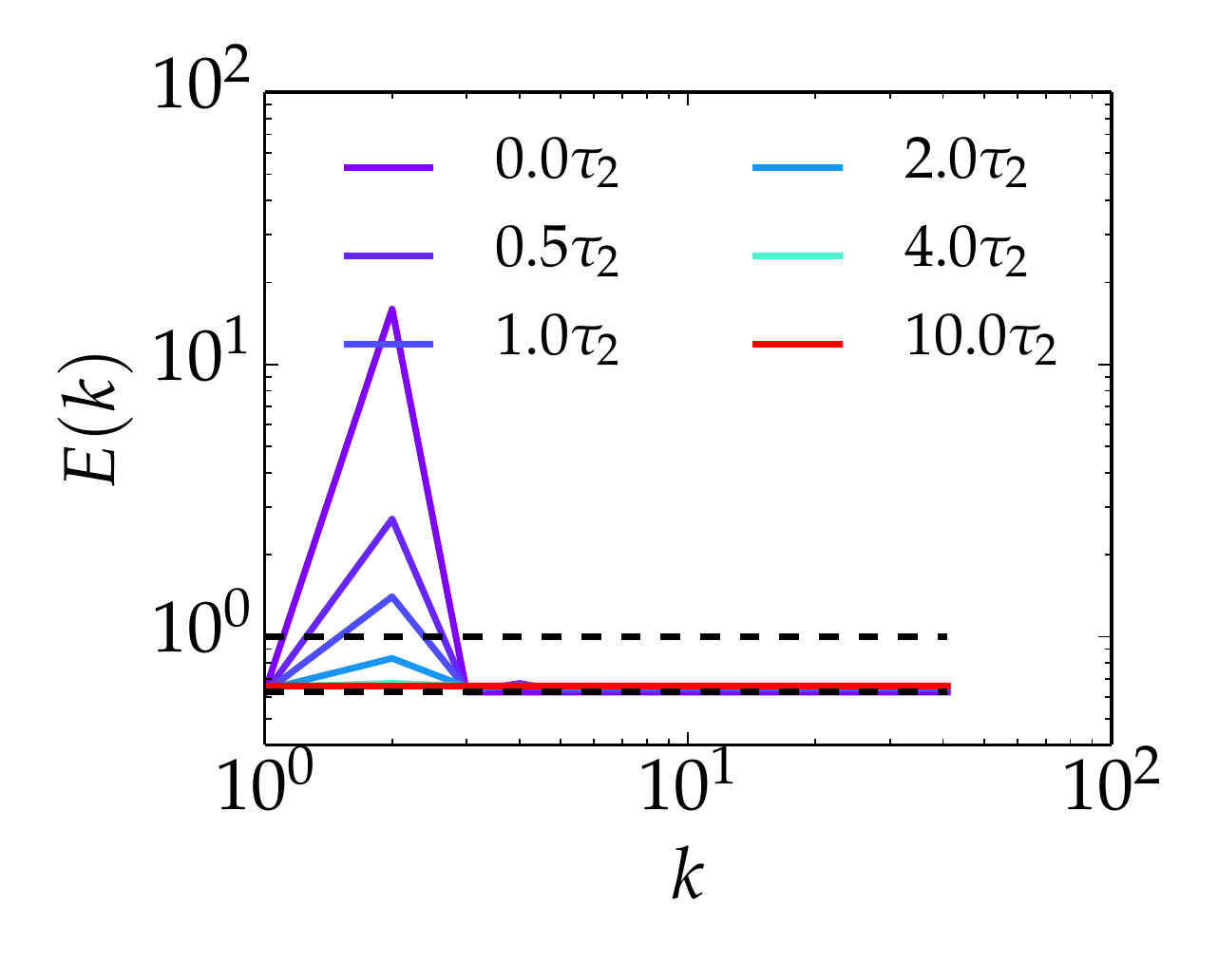}\\
		\end{minipage}
	\caption{The left panel compares the mean-field (\ref{eq:MFoptimal}) and the iterative determination of  the optimal control $\sigma_{opt,2}$, that determines the  stationary response to a single-mode disturbance of the mode $k=2$ out of $K=41$ active modes, for initial disturbances $b_2 = \beta/1.1$ (\textcolor{NavyBlue}{$\bullet$}) , $\beta/2$ (\textcolor{Cyan}{$\bullet$}) and $\beta/16$ (\textcolor{Brown}{$\bullet$}).  The upper right panel shows the time evolution of the total energy as determined by the numerical optimization.  The lower right panel  shows the corresponding evolution of the normalized energy spectrum $E(k) = \beta/b_k$ from the highest initial perturbation $b_2 = \beta/16$ towards a ``wrong'' equipartition state.  The descent optimizes the cost between $0$ and $t = 10 \tau_2$, the initial total energy is $1$, and convergence is declared when the amplitude of the cost gradient becomes smaller than $g_{tol}= 10^{-6}$.}
	\label{fig:dissip-optimalcontrol}
	\end{figure}

	\paragraph{Mean-field behavior.}
	As a validation of the algorithm, we compute the  stationary (shadow) response $\tilde b(\cdot |\infty)$ to the initial perturbation of a  single-mode ${k}$ disturbed away from equipartition, say
	 \begin{equation}
		\bb_{0,l} =  b_{k,0} \text{~if $ l = k$, and~}  \dfrac{K\beta-b_k}{K-1} \text{~~otherwise}.
		\label{eq:single-mode}
	\end{equation}
In the limit of a very large number of modes $ K \gg 1$, the system becomes one-dimensional,  as all the modes but one are thermalized to $ \beta = K/E$ (see \ref{sec:MeanFieldApprox}). The optimal control is then one-dimensional, and hence its solution is simply deduced  from  the one-dimensional ordinary differential equation~: 
	\begin{equation}
		\sigma_{opt,k}(t) = \dfrac{\dot b_k}{b_k},   \text{~~where~}  \dot b_k=  \sqrt 2 \dfrac{ \beta }{\tau_k} \left(\dfrac{b_k}{\beta}\right)^{1/2}  \left(1-\dfrac{b_k}{\beta}\right) \text{ ~and~} \tau_k = k^{-1}E^{-1/2}.
		\label{eq:MFoptimal}
	\end{equation}
	Figure  \ref{fig:dissip-optimalcontrol} shows the optimal control that the  descent algorithm converges to, when out of $ K=41$ modes the mode  $k=2$ is disturbed according to (\ref{eq:single-mode}).  Good agreement is found with the mean-field solution (\ref{eq:MFoptimal}) for the lowest perturbation $ b_k = \beta/1.1$ (see the left panel of Figure \ref{fig:dissip-optimalcontrol}).  The energy is conserved along the shadow path, and at final time the energy spectrum cannot be distinguished from the equipartition state $1/\beta$ (not shown). This illustrates the rational behavior of the algorithm.
	
	\paragraph{Non mean-field behavior.}
	As we increase the amplitude of the initial perturbation   from $\beta/ 2$ to $\beta/16$, deviations from the mean-field become more and more pronounced. It is  interesting to remark that in this non-mean field regime, the total energy is not preserved along the shadow path, and therefore neither along the optimal path.  As a consequence, the path reaches a  wrong state of equipartition, whose total energy is lower than the initial one. The failure is particularly spectacular for the largest perturbation. 
	 The non-conservation of the energy feature is not a failure of the algorithm, but an unwanted consequence of the  degeneracy of the lack-of-fit potential $\mU[\bb]$ defined in (\ref{eq:TBHLofLagrangian}). Fortunately, this defect can be fixed as follows.  
	\subsection{Constrained minimization and Augmented Lagrangian formalism.}
	Several strategies can be used to  tie the final equipartition state  to the initial value of the ensemble energy, either at the level of the trial densities  or at the level of the optimization problem. We decide for the latter option. We redefine the optimal control as the solution to the following \emph{constrained} minimization problem, where the statistical conservation of the energy  is imposed by a  constraint $\mC(\bb,\bsigma)$ that depends on the state and the control :
	\begin{equation}
		\underset{[\bsigma]}{\inf} \int_0^t\d s \;  L_{lof}\left(\bb(s),\bsigma(s)\right) \text{~subject to ~} %
		\begin{cases}
			& \bb[t=0] = \bb_0,	\\
			&  \dot \bb = \bphi(\bb,\bsigma) \text{~with ~}  \phi_l(\bb,\bsigma) = b_l \sigma_l,	\\
			&  \text{and~} 0 = \dot E = \mC(\bb,\bsigma)=- \sum_{l=1}^K \sigma_l/b_l.
		\end{cases} 
	\label{eq:optimalmodif}
	\end{equation} 
Robust numerical algorithms  are documented to solve such globally constrained optimization problems, one example being the  augmented Lagrangian method (see \cite[Chapters 12 and 17]{nocedal_numerical_2006} and references therein). To enforce energy conservation, we use a time-dependent Lagrange multuplier $\lambda(s)$ and a scalar penalty factor $ \mu$  in the  augmented  objective function~:
		\begin{equation}
			\begin{split}
			&\tilde \mJ_{\lambda,\mu}[[\bsigma],t] = \int_0^t\d s \;\left(  \dot \bb(s) \cdot \bpi (s) - \mH_{P,\lambda,\mu}(\bb(s),\bpi(s),
			\bsigma(s))\right),\\
		& \text{where ~} \mH_{P,\lambda,\mu} = \bphi \cdot \bpi - L_{lof}(\bb,\bsigma)  + \lambda\mC(\bb,\bsigma) - \dfrac{1}{2\mu}\mC^2(\bb,\bsigma)
			\end{split}
			\label{eq:extentedlag_withweight}
		\end{equation}
now defines the augmented Hamilton-Pontryagin function.
The descent algorithm and the updating scheme for $\lambda $ and $\mu$ are described in  \ref{sec:DiscretePontryagin} (Algorithm \ref{algo:lancelot}). The convergence is declared when both the gradient of the objective cost and the constraint norm become smaller than pre-defined tresholds, say  $g_{tol}$ and $c_{tol}$.

\begin{figure}
		\begin{minipage}{0.59\textwidth}
			\centering
			\includegraphics[width=\textwidth]{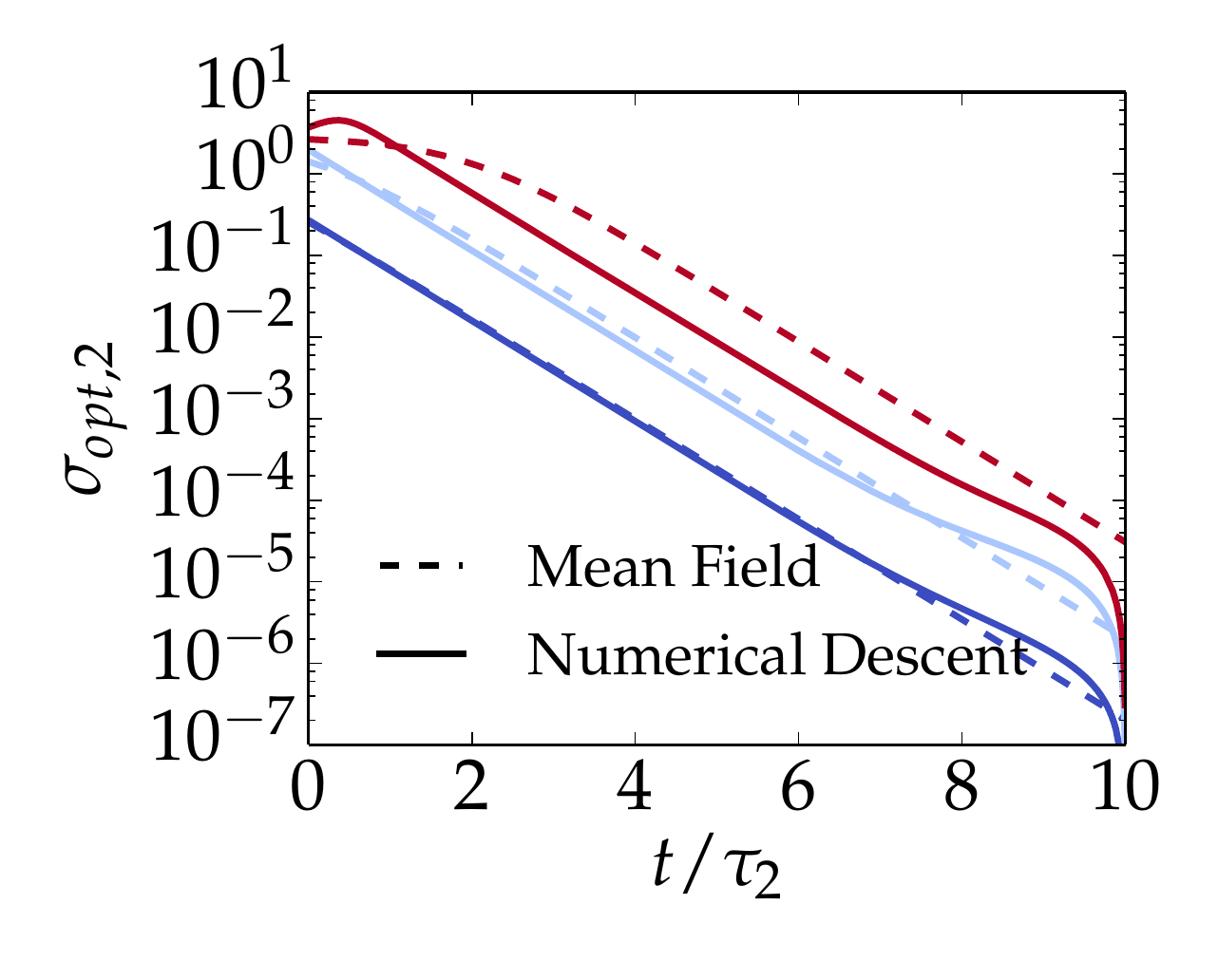}\\
		\end{minipage}
		\begin{minipage}{0.4\textwidth}
			\centering
			\includegraphics[width=\textwidth]{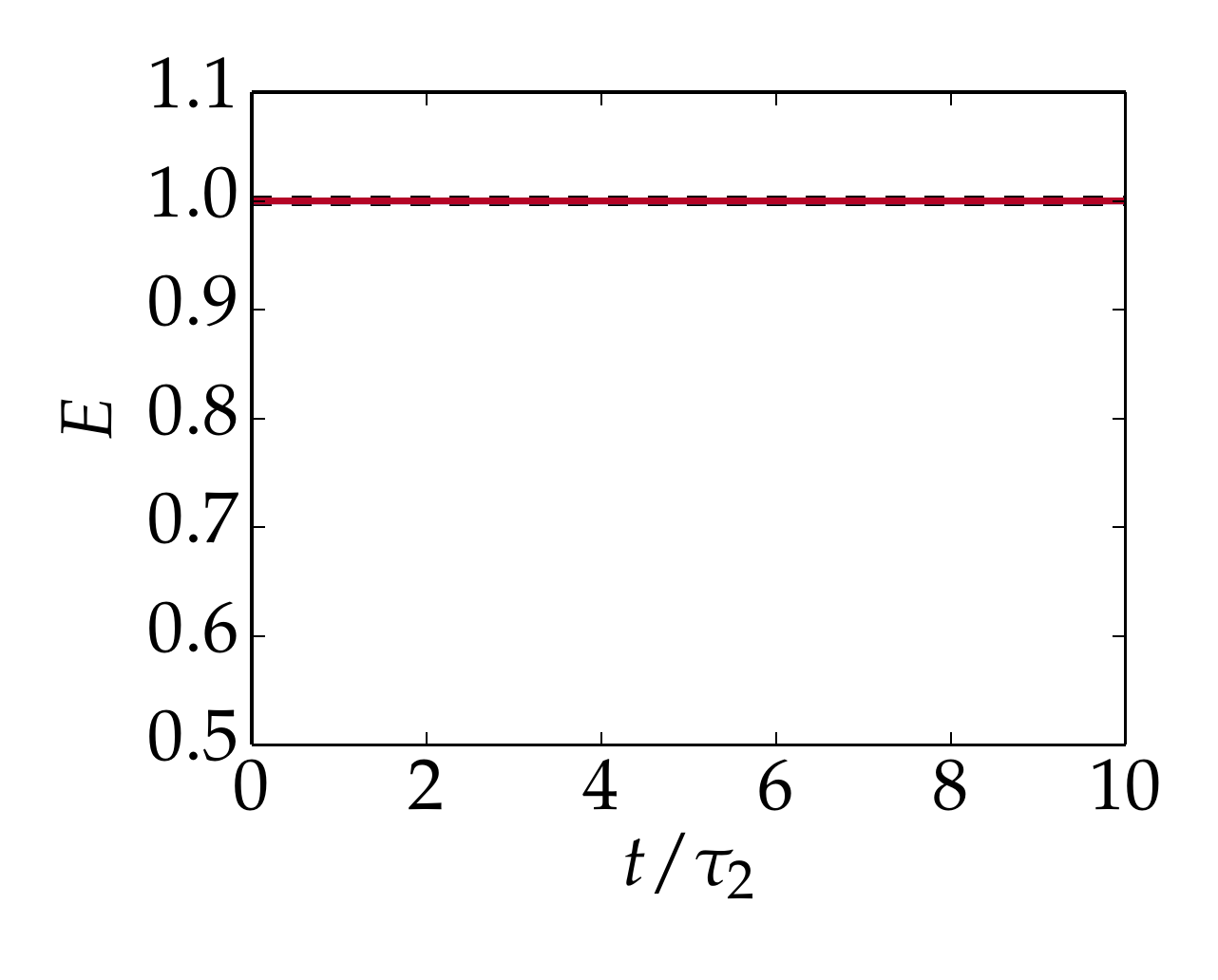}
			\includegraphics[width=\textwidth]{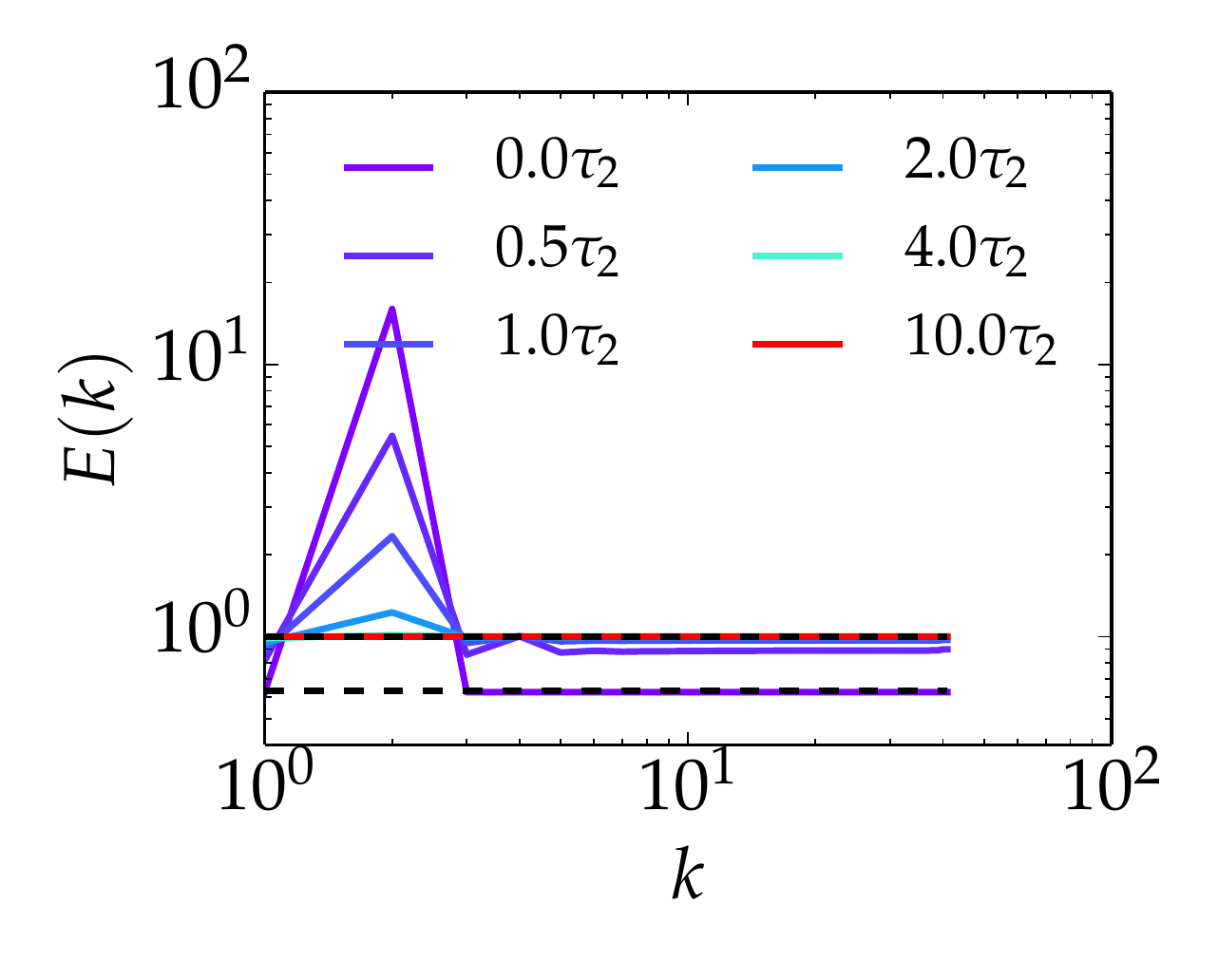}\\
		\end{minipage}
	\caption{Same as Figure \ref{fig:dissip-optimalcontrol}, but the iterative determination now incorporates  the conservation of the energy through the augmented Lagrangian method.  The lower right panel now shows the  stationary evolution of the the normalized energy spectrum $E(k) = \beta/b_k$  from the highest initial perturbation $b_2 = \beta/16$ towards the correct equipartition state. Convergence is declared when the amplitude of the augmented cost gradient becomes smaller than $g_{tol}= 10^{-6}$, and the constraint norm is smaller than $c_{tol} = 10^{-4}$.}
		\label{fig:cons-optimalcontrol}
	\end{figure}

Figure  \ref{fig:cons-optimalcontrol} illustrates the algorithm's consistency for the single-mode disturbance scenario, and shows the convergence towards the correct equipartition state. The departure from the mean-field prediction for the largest perturbations is due to the mean-field assumption that all the undisturbed modes remain exactly in equipartition.
\subsection{The optimal response algorithm.}
\label{sec:Pontryalgo}
The optimal response algorithm (Algorithm \ref{algo:optimalresponse}) sequentially pieces together the augmented Lagrangian method and the non-linear conjugate gradient descent to compute the optimal response with the desired accuracies $g_{tol}$ and $c_{tol}$ up to time $t$.  The process essentially consists in first  approximating the envelope with loose convergence criteria, and then in refining the envelope up to the prescribed accuracy. The second step can be done in parallel. 
 \begin{algorithm}[h]
\caption{The optimal response algorithm}\label{algo:optimalresponse}
\begin{algorithmic}[1]
\STATE Set a sequence $T_0=0 <T_1 <\dots <T_k = t$ of final times. The $T_i$'s need not be linearly spaced.
\STATE For each $T_i$, compute the shadow response $\tilde b(\cdot |T_i)$ using the augmented Lagrangian algorithm up to relaxed convergence criterion $g_{tol}^\prime > g_{tol}$ and $c_{tol}^\prime > c_{tol}$ . As a first guess for the initial values of both the  control and the augmentation weights, take the converged values corresponding to time $T_{i-1}$.
\STATE  For each $T_i$, use the previously ``loosely'' converged controls and augmentation weights to initialize the augmented-descent algorithm and reach the prescribed accuracies $g_{tol}$ and $c_{tol}$.
\STATE Interpolate the final time values of the final states to obtain the optimal response.
\end{algorithmic}
\end{algorithm}
%
%
 \newenvironment{decaywithspecinset}[2]
 {
 \begin{minipage}{0.49\textwidth}
	\begin{tikzpicture}
	 \node[anchor=south west,inner sep=0] (image) at (0.5,0) {\includegraphics[width=0.98\textwidth,trim=1.7cm 0cm 0.5cm 0cm,clip]{#1-Energy}};
 	 \begin{scope}[x={(image.south east)},y={(image.north west)}]
	 	 \node [anchor=east,rotate=90] at (0,0.7) {#2};
		\node[anchor=south west,inner sep=0,fill opacity=1] (image) at (0.45,0.65) {\includegraphics[width=0.5\textwidth,trim=1.9cm 1.7cm 0.8cm 0.75cm,clip]{#1-EnergySpec}};
		\node[anchor=south,inner sep=0,fill opacity=1] (image) at (0.85,0.72){\small $k$};
		\node[anchor=east,inner sep=0,fill opacity=1,rotate=90] (image) at (0.43,0.92){\small $E_k$};
	\end{scope}
	\end{tikzpicture}
\end{minipage}
	};
 \newenvironment{decaywithspecinsetb}[3]
 {
 \begin{minipage}{0.49\textwidth}
	\begin{tikzpicture}
	 \node[anchor=south west,inner sep=0] (image) at (0.5,0) {\includegraphics[width=0.98\textwidth,trim=1.7cm 1.7cm 0.5cm 0cm,clip]{#1-Energy}};
 	 \begin{scope}[x={(image.south east)},y={(image.north west)}]
		  \node [anchor=south east,rotate=90] at (0,0.6) {#2};
	 	 \node [anchor=north,rotate=0] at (0.6,0) {#3};
		\node[anchor=south west,inner sep=0,fill opacity=1] (image) at (0.4,0.65) {\includegraphics[width=0.55\textwidth,trim=1.9cm 1.7cm 0.8cm 0.75cm,clip]{#1-EnergySpec}};
		\node[anchor=south,inner sep=0,fill opacity=1] (image) at (0.85,0.72){\small $k$};
		\node[anchor=east,inner sep=0,fill opacity=1,rotate=90] (image) at (0.37,1){\small $E_k$};
	\end{scope}
	\end{tikzpicture}
\end{minipage}
	};

  \newenvironment{decaywithspecinsetlowE}[3]
 {
 \begin{minipage}{0.49\textwidth}
	\begin{tikzpicture}
	 \node[anchor=south west,inner sep=0] (image) at (0.5,0) {\includegraphics[width=0.98\textwidth,trim=1.7cm 1.7cm 0.5cm 0cm,clip]{#1-Energy}};
 	 \begin{scope}[x={(image.south east)},y={(image.north west)}]
		  \node [anchor=south east,rotate=90] at (0,0.6) {#2};
	 	 \node [anchor=north,rotate=0] at (0.6,0) {#3};
		\node[anchor=south west,inner sep=0,fill opacity=1] (image) at (0.42,0.48) {\includegraphics[width=0.52\textwidth,trim=1.9cm 1.7cm 0.8cm 0.75cm,clip]{#1-EnergySpec}};
		\node[anchor=south,inner sep=0,fill opacity=1] (image) at (0.85,0.55){\small $k$};
		\node[anchor=east,inner sep=0,fill opacity=1,rotate=90] (image) at (0.4,0.82){\small $E_k$};
	\end{scope}
	\end{tikzpicture}
\end{minipage}
	};

 \newenvironment{decaywithspecinsethighE}[3]
 {
 \begin{minipage}{0.49\textwidth}
	\begin{tikzpicture}
	 \node[anchor=south west,inner sep=0] (image) at (0.5,0) {\includegraphics[width=0.98\textwidth,trim=1.7cm 1.7cm 0.5cm 0cm,clip]{#1-Energy}};
 	 \begin{scope}[x={(image.south east)},y={(image.north west)}]
		  \node [anchor=south east,rotate=90] at (0,0.6) {#2};
	 	 \node [anchor=north,rotate=0] at (0.6,0) {#3};
		\node[anchor=south west,inner sep=0,fill opacity=1] (image) at (0.21,0.11) {\includegraphics[width=0.4\textwidth,trim=1.9cm 1.7cm 0.8cm 0.75cm,clip]{#1-EnergySpec}};
	\end{scope}
	\end{tikzpicture}
\end{minipage}
	};
\section{Numerical experiments.}
	We now use the optimal response algorithm to examine the predictive skill of the optimal response framework for Burgers dynamics, and compare it to both direct numerical simulations (DNS)  and F/D type predictions. More specifically, we study the non-equilibrium response to  two specific types of perturbations \emph{(i)} a ``single-mode disturbance'' away from equipartition, and \emph{(ii)} a ``many-mode disturbance''. For both scenarios, we find that the optimal and the F/D predictions have a comparable range of relevance. 
	\subsection{Single-Mode vs Many-Mode disturbances.}
		In the  single-mode perturbation scenario, a single mode $ k$ carries most of the disturbance  away from equipartition. This is the case previously described by Equation (\ref{eq:single-mode_Recall}), which we here define in terms of the disturbance amplitude $\Delta_0$ as~:
	\begin{equation}
		\bb_{0,l} =  \beta / \Delta_0 \text{~if $ l = k$, and~}  \dfrac{K-1}{K-\Delta_0} \beta \text{~~otherwise}.
		\label{eq:single-mode_Recall}
	\end{equation}
		In the  many-mode perturbation scenario, the $k-1$ modes graver than $k$ carry the same amount of disturbance  away from equipartition.  The initial ensemble is then taken as~:
	\begin{equation}
		\bb_{0,l} =  \beta / \Delta_0 \text{~if $ l <  k$, and~}  \dfrac{K-k+1}{K-k\Delta_0} \beta \text{~~otherwise}.
		\label{eq:single-mode_Recall}
	\end{equation}
In both cases, the (average) equipartition energy contained in each shell is $E_{eq}=1/\beta$, so that the total energy is $E = K/\beta$. In our numerics, we systematically set the total energy to $1$.   
The amplitude $\Delta_0$ represents the excess energy in the disturbed modes $k$~: at initial time, they each  have the same energy $E_k = \langle |v_k|^2 \rangle = \Delta_0 E_{eq}$.
	\subsection{Computing Averages.}
\begin{figure}
		\includegraphics[width=0.49\textwidth]{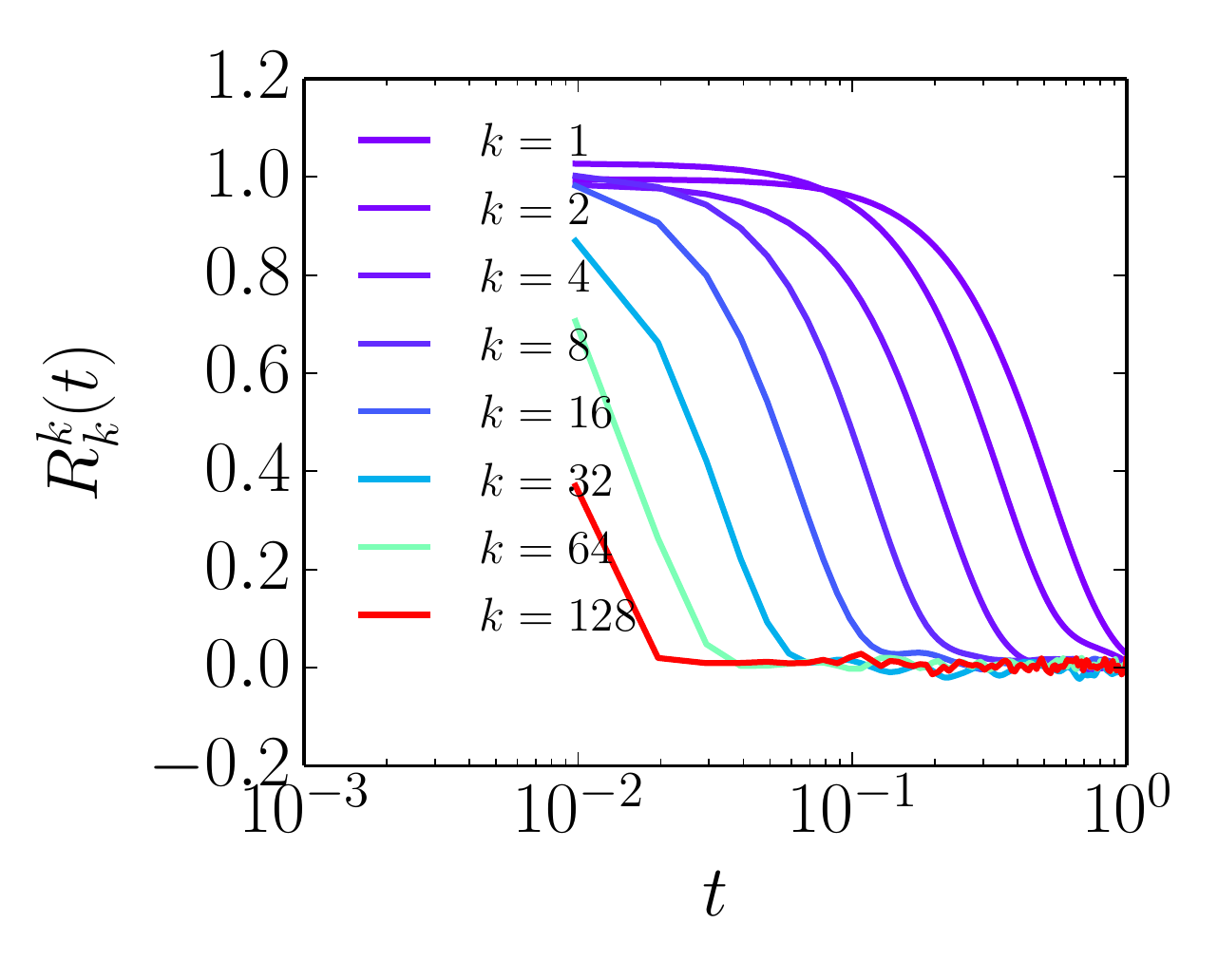}
		\includegraphics[width=0.49\textwidth]{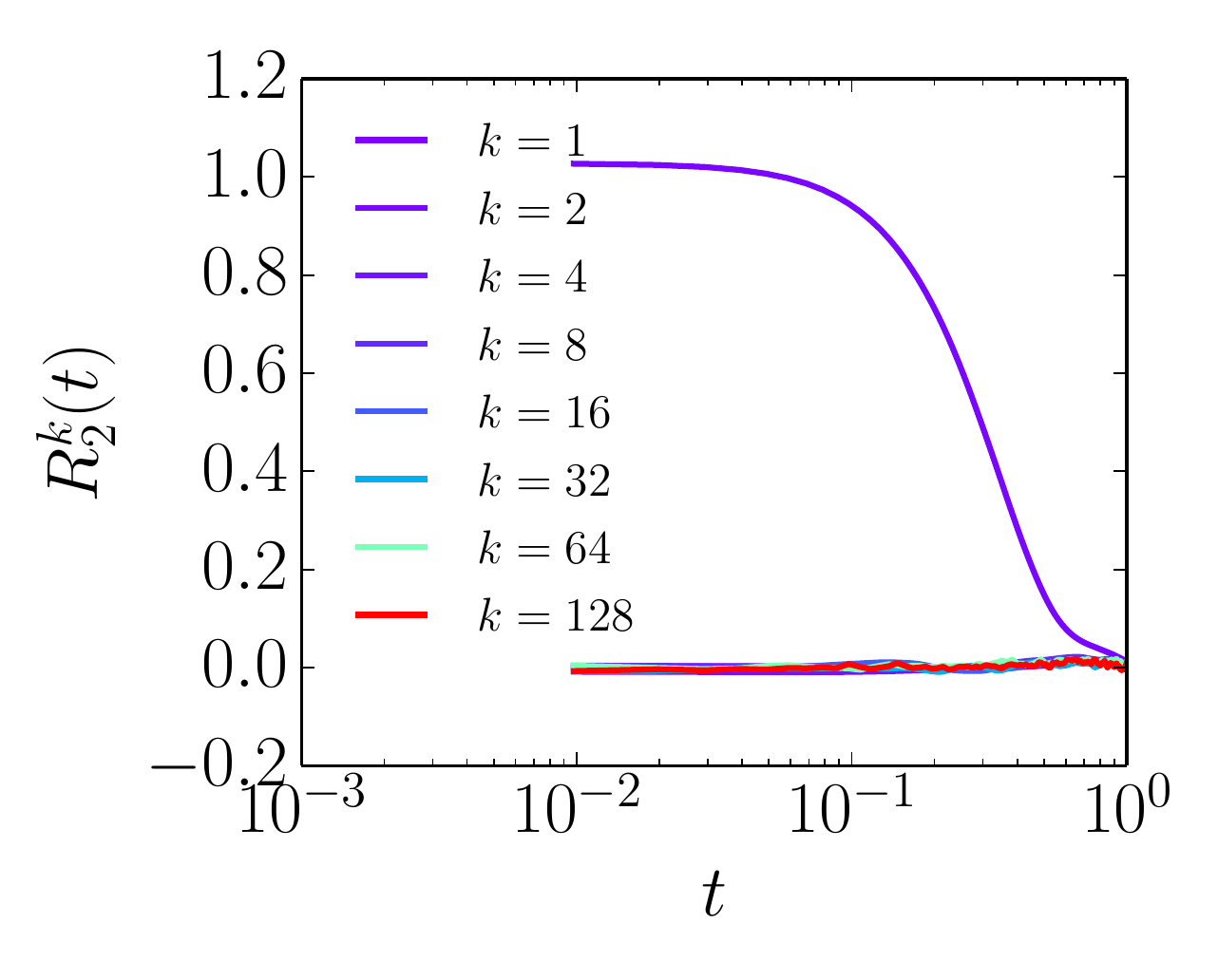}
	\caption{Short-time behavior of the diagonal elements (left) and the two-mode correlation $(2,k)$ (right) of the  response function, computed by averaging over 20,000 realizations of  512-resolved DNS. }
	\label{fig:ResponseFunction}
	\end{figure}

		For both scenarios (single-mode and many-mode disturbances), we estimate the relaxation of the energy spectrum in three different ways : using DNS to estimate the true ensemble averages, using the F/D  estimate (\ref{eq:FDestimateEnergy}), and using the optimal response algorithm for the optimal estimate (\ref{eq:optimalestimateEnergy}).
		\paragraph{DNS.}
		 True non-equilibrium averages are estimated by making ensemble averages from individual realizations of the relaxation, computed with DNS.  The TBH dynamics (\ref{eq:tbh}) is integrated in time with  a standard $4^{th}$-order Runge- Kutta algorithm. The non-linear terms are estimated with a pseudo-spectral method,  that uses the 2/3-rule dealiasing \cite{orszag_elimination_1971}.  The results that we report here correspond to  DNS with  spatial resolutions of $N = 128$ and $N = 512$ grid points. The corresponding cutoffs in Fourier space are then exactly  $K =42$ and $K=170$. The time-steps are  taken as $\delta t  = 3.9 \times 10^{-3}$ and $9.8 \times 10^{-4}$, and guarantee an accurate conservation of the energy. Averages are taken over 20,000 realizations.
		\paragraph{F/D estimate.}
		To estimate the energy spectrum relaxation with the F/D relations, we need to  determine numerically the response functions $R_k^l$. This is done by performing averages of DNS realizations sampled from equipartition, with parameters similar to those defined in the previous paragraph.  To improve the statistical accuracy, we use the symmetric part of the correlation functions $\langle |v_k(t)|^2| v_l(0)|^2\rangle_{DNS}$ to determine the response function. This is justified, provided that the underlying statistics are indeed stationary. 	Figure \ref{fig:ResponseFunction} shows the short time-behavior of the so-determined response functions for a $512$ resolution.

		\paragraph{Optimal Response.}
	The optimal estimates of the non-equilibrium averages are obtained from the optimal response algorithm. The numerical precisions are set to $g_{tol} = 10^{-6}$ and $c_{tol}=10^{-4}$. The time-steps are $\delta t^\prime = 1.2 \times 10^{-2}$ for the 128 case, and $\delta t ^\prime= 7.8 \times 10^{-4}$ for the 512 case. The final times $T$ at which the shadow responses are computed are taken to be logarithmically spaced between $T_1= 4 \delta t$ and $T_f = 5.$ For the 128 case, we refine the short-time computation  of the optimal enveloppe by determining additional shadow responses between $T = 4 \delta t^\prime$ and $T =0.1$. For those additional points, we use the value $\delta t ^\prime$ for the time-steps of the shadow responses.

	\subsection{Results.}
\begin{figure}
		\begin{minipage}{0.85\textwidth}
		\decaywithspecinsetb{OneMode_128_E1}{\large $E_k$}{}
		\decaywithspecinsetb{OneMode_128_E2}{}{}\\
		\decaywithspecinsetb{OneMode_128_E4}{\large $E_k$}{\large $t/\tau_k$}
		\decaywithspecinsetb{OneMode_128_E8}{}{\large $t/\tau_k$}
		\end{minipage}
		\begin{minipage}{0.14\textwidth}
		\includegraphics[width=\textwidth,trim=0cm 0cm 0cm 0.2cm,clip]{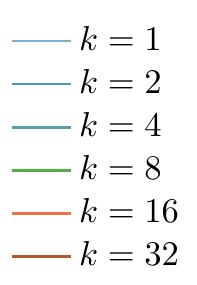}
		\end{minipage}
	\caption{The figures show the decay of the most perturbed mode $k$ back to equipartition after  single-mode disturbance, for $\Delta_0 = 1.1,2,4,8 $ (from left to right and top to bottom). The energies are normalized to their equipartition value $1/\beta$. The DNS resolution is $N=128$. The colors code the wavenumber of the disturbed mode. The  symbols code the averages : DNS (solid line), F/D (diamonds) and optimal (circles). The insets show the corresponding energy spectra, which is perturbed at initial time and undistinguishable from equipartition at final time.}
	\label{fig:128-OneMode}
	\end{figure}

\begin{figure}[h]
		\begin{minipage}{0.87\textwidth}
		\decaywithspecinsetb{OneMode_512_E2}{\large $E_k$}{\large $t/\tau_k$}
		\decaywithspecinsetb{OneMode_512_E8}{}{\large $t/\tau_k$}
		\end{minipage}
		\begin{minipage}{0.12\textwidth}
		\includegraphics[width=\textwidth,trim=0.1cm 0cm 0cm 0.2cm,clip]{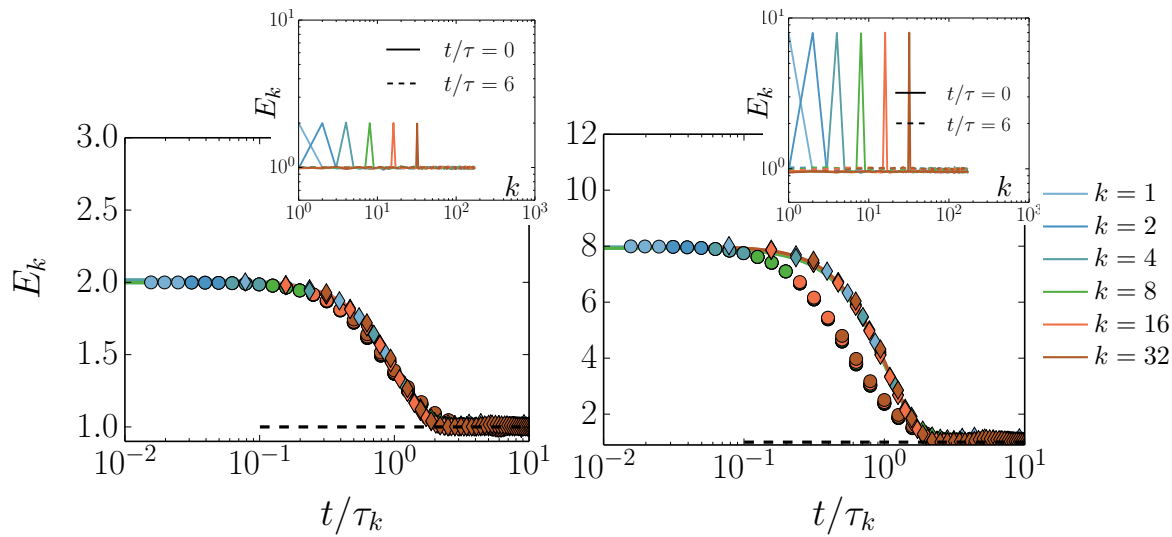}
		\end{minipage}
	\caption{Same as Figure \ref{fig:128-OneMode}, for a DNS resolution $N =512$. Only the respones to the single-mode disturbances $\Delta_0 = 2$ (left) and $\Delta_0=4$ (right).}
		\label{fig:512-OneMode}
	\end{figure}

	\paragraph{Single-mode disturbances.}
	We show the responses to single-mode perturbations for both low and high modes, namely $k=1,2,4,8,16$ and $32$, and using $\Delta_0 = 1.1,2,4 $ and $8 $ for the amplitude of the initial disturbances. The numerical outcomes are summarized in Figures (\ref{fig:128-OneMode})  and (\ref{fig:512-OneMode}), which display the decay towards equipartition of the three estimates for the non-equilibrium averages of the disturbed energy  at mode $k$. We use as the natural time-scale the mean-field time $\tau_k = k^{-1}E^{-1/2}$ previously encountered in Equation (\ref{eq:MFoptimal}).
For the smallest perturbations $ \Delta_0 = 1.1$ and $2$, the three averages (true, F/D and optimal) are indistinguishable from one another. This is visible for both the $N=128$ and $N=512$ cases. The collapse is perfect in all of the  three stages of the energy relaxation : the initial stage up to
$ t \approx 0.2\tau_k$ where the energy remains essentially constant, the intermediate relaxation stage 
$0.2 \tau_k \lesssim t \lesssim \tau_k$ and the final equilibrium stage after $\tau_k$. Discrepancies start to be noticeable for the two highest initial disturbances. The optimal responses capture the natural time scaling and the collapse of the relaxation profiles when the time is normalized by $\tau_k$. However, the initial stage of zero dissipation of energy ends too soon, and the intermediate stage is slightly longer than in the DNS. This feature is particularly visible for the 512 runs  (see Figure \ref{fig:512-OneMode}, right panel).

\begin{figure}
		\begin{minipage}{0.87\textwidth}
		\decaywithspecinsetlowE{ManyMode_512_E2_K4}{\large $E_k$}{}
		\decaywithspecinsetlowE{ManyMode_512_E2_K16}{}{}\\
		\decaywithspecinsethighE{ManyMode_512_E8_K4}{\large $E_k$}{\large $t/\tau_k$}
		\decaywithspecinsethighE{ManyMode_512_E8_K16}{}{\large $t/\tau_k$}
		\end{minipage}
		\begin{minipage}{0.12\textwidth}
		\includegraphics[width=\textwidth,trim=0cm 0cm 0cm 0cm,clip]{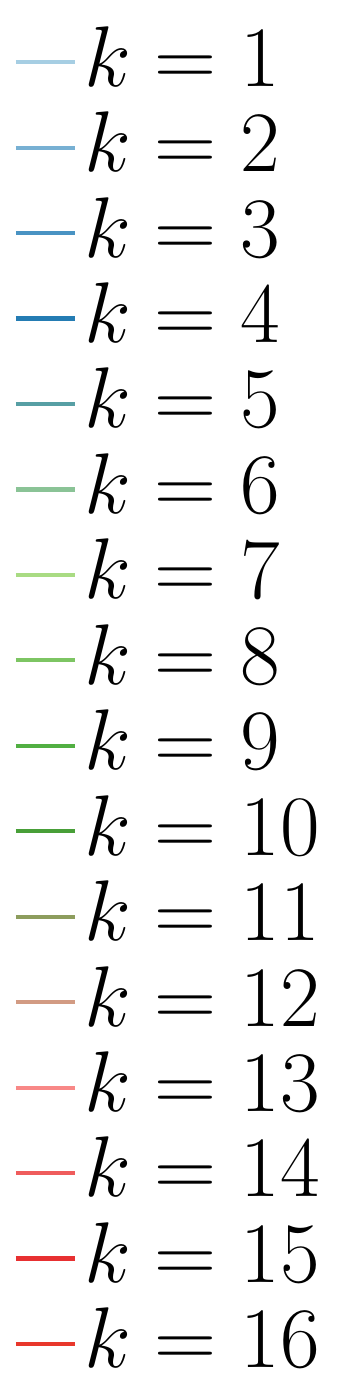}
		\end{minipage}
	\caption{The figures show the decay of the $k-1$ most perturbed modes back to equipartition after  a many-mode disturbance, for $\Delta_0 =2$ (top) and $8 $ (bottom) and $k=2$ (left) and $16$ (right) . The energies are normalized to their equipartition value $1/\beta$. The DNS resolution is $N=512$. The colors code the wavenumber of the disturbed modes. The  symbols code the averages : DNS (solid line), F/D (diamonds) and optimal (circles). The insets show the corresponding energy spectra, which is perturbed at initial time and undistinguishable from equipartition at final time.}
	\label{fig:512-ManyMode}
	\end{figure}

	\paragraph{Many-mode disturbances.}
	The same qualitative features are observed for the many-mode disturbance scenario, as displayed on Figure 	\ref{fig:512-ManyMode}. For the small amplitude perturbations $\Delta_0 =2$ displayed on the top panel, the agreement between the three types of averages is excellent, whether the number of disturbed modes is small or large.
	Discrepancies are apparent for larger initial perturbations (bottom panel), for which the ``undisturbed'' higher modes are initially far away from their thermalized values. For the case where only the first three modes are disturbed, we observe as in the single-mode case that the optimal response has a good qualitative behavior but starts to decay too fast compared to DNS. In that case, the F/D estimate apparently gives a correct prediction of the DNS averages.
	When a significant number of modes are taken away from equipartition, however, both the F/D and the optimal responses break down. It is interesting to observe that in that case, the optimal responses do not collapse any longer with the time scale $\tau_k$, and nor do the DNS. The qualitative behavior of the optimal response is surprising. It predicts an energy transfer from the low modes to the high modes during the initial stage, resulting in a too fast decay of the high modes and a too slow decay  for the low modes, characterized by an initial bump in the energy profile.
We therefore conclude that the optimal response is accurate for perturbations of moderate amplitudes, in which case it displays a mean-field behavior and is indistinguishable from the true DNS. For higher perturbations, the skill of the optimal response theory breaks down, as seemingly non-physical inverse energy transfers are predicted. It appears that the range of applicability of the optimal responses is slightly more restricted than that of the F/D approach.

\section{Conclusion.}
In this paper we have shown that the best-fit dynamical optimization recently exposed in a series of papers provides a  fully predictive theory, as one can use standard  optimization algorithms to determine the optimal response. The optimal response algorithm described in  Section 3 achieves such a task. 
We were able to test the skill of the best-fit theory  \emph{per se}, without relying to further approximations (mean-field, perturbation expansions) to solve the underlying Hamilton-Jacobi equation.
In the near-equilibrium regime, the optimal estimates are consistent with DNS, and share a similar range of applicability as  linear F/D estimates -- although perhaps slightly more restricted.  In contrast to the latter, we emphasize that the optimal responses are computed self-consistenly and independently from DNS.  This makes the optimal response an approximate but fully predictive theory.

From a more conceptual point of view, the use of the optimal response algorithm beyond the mean-field regime revealed a flaw of the original dynamical optimization, namely that the optimal responses  fail to conserve energy in general. The defect is an undesired consequence of the degeneracy of the lack-of-fit potential, but can be easily fixed by imposing energy conservation in the optimization principle. Perhaps,  the origin of the degeneracy relates to the single-time nature of the theory. 

We have restricted our exposition to simple quasi-Gibbsian trial densities. In principle, patience is the only virtue required to compute the lack-of-fit Hamiltonian associated to more accurate trial densities. The optimal response algorithm provides a a systematic way to compute the associated best-fit response and can in principle be used to investigate more refined Ans\"atze that include non-Gibbsian parts. We hope that those could prove efficient to model the decay of far-from-equilibrium disturbances. Let us however  observe that the algorithm performs a multi-layer optimization descent on a high dimensional space. In the quasi-Gibbsian case, the underlying assumption is that the many-mode correlations need not be modeled to give accurate estimates, so that the number of variables in the optimization problem scales linearly with the size of the problem.  This hypothesis will break down further away from equilibrium.  In order to optimize over many-mode correlations, an appropriate modeling effort will therefore be required at the level of the trial densities themselves, in order to make the numerical optimization  computationally tractable.


\appendix
\section{The mean-field approximation to the lack-of-fit Hamiltonian}
\label{sec:MeanFieldApprox}
The optimal response for a single-mode perturbation $k$ away from equipartition can be obtained via a mean-field argument,
that consists in putting $b_l(t) = \beta$ for the non-perturbed modes in the lack-of-fit  Hamiltonian (\ref{eq:TBHLofHamiltonian})
The resulting reduced lack-of-fit Hamiltonian now depends on a single pair of conjugate variables $(b_k,\pi_k)$, which we write suggestively  as follows~:
\begin{equation}
	\begin{split}
		&\mH_{lof}(b_k,\pi_k) = \dfrac{\pi_k^2 b_k^2}{2} - \mU[b_k] ,  \text{~with~~} \mU[b_k] = C_1 \left(\dfrac{1}{b_k} -\dfrac{1}{\beta}\right)+ C_2\left( \dfrac{b_k}{\beta^2} -\dfrac{1}{\beta}\right)  \\
		& \text{and}~ C_1 = k^2K\left( 1 - \dfrac{k+7}{2K} \right) ~ , ~ C_2 = k^2K\left( 1 - \dfrac{k+3}{2K} \right).
	\end{split}
	\label{eq:TBHLofHamiltonian_reduced}
\end{equation}
For $k/K \ll 1$ , which corresponds to a large-scale perturbation, we can recast the lack-of-fit potential as 
\begin{equation}
	\begin{split}
		&\mU[b_k] = \dfrac{\beta}{\tau_k^2 b_k} \left( 1 - \dfrac{b_k}{\beta}\right)^2  \text{~with~} \tau_k = k^{-1} E^{-1/2} \text{~and $E$ the total energy.}\\ 
	\end{split}
	\label{eq:TBHLofClosurePotential_reduced}
\end{equation}
The stationary shadow path is then defined by the equality $\mH_{lof}(b_k,\pi_k)= \mH_{lof}(\beta,0) =0$, which leads directly to (\ref{eq:MFoptimal}).
In the mean-field approximation, the non-perturbed modes $l \neq k$ act as a thermal bath that sinks the excess energy induced by the initial perturbation. As such, the total energy is not conserved.

%
\section{Discrete formulation of the optimal response algorithm.}
\label{sec:DiscretePontryagin}
	In this appendix, we expose the details of the optimal response algorithm. The practical implementation relies on a discrete approximation, which is performed directly at the level of the objective cost. Standard  algorithms can then be used to solve the discrete optimization problems.
	\subsection{Discretizing the optimization problem.}
	In this subsection, we define a discrete version of the minimization problem (\ref{eq:optimalmodif}). We define a discrete counterpart to the augmented objective cost, that involve a discrete augmented lack-of-fit Pontryagin Hamiltonian.

	\paragraph{Discrete Notations.}
	We write $t$ the final time involved in  (\ref{eq:optimalmodif}).  We define the time step $\Dt = t/\nt$, and discretize the time interval $[0;t[ $ into the $\nt$ time intervals $[t_i;t_{i+1}[$, delimited by the discrete times $t_i = i \Dt$.
	We write   $\Bs_i = \Bs(t_i)$, $\bpi_i = \bpi(t_i)$ and  $\bsigma_i = \bsigma(t_i)$,  the values of the state, co-state and control variables at the discrete times $t_i$. The notation $\Bs$, without subscript, denotes  $\Bs=(\Bs_i)_{0\le i \le \nt}$. Recall that the $(\Bs_i)$'s are $\mathbb R^K$ vectors, so that for example $s_{i,l}$ denotes the $l^{th} $ component of the state variable at time $t_i$, and that we use the notation $\ba_i \cdot \bb_i = \sum_{l=1}^K a_{i,l}b_{i,l}$.
	Note that we here prefer to use the symbol  $\Bs$ to denote the state variable, so that the latter is not necessarily taken to be the inverse temperature $\bb$, as was the case in Section (\ref{sec:Pontryagin}). 
	
	We  use the trapezoidal rule to estimate the time integrals. The discrete counterpart to the minimization problem (\ref{eq:optimalmodif}) is then 
		\begin{equation}
		\underset{\bsigma}{\inf}~  \Dt \sum_{i=0}^{\nt-1}  \dfrac{L_i + L_{i+1}}{2} \text{~subject to ~} %
		\begin{cases}
			& \Bs[t=0] = \Bs_0,	\\
			&  \Bs_{i+1} - \Bs_{i} =  h \,\bphi(\Bs_{i},\bsigma_{i}) = h \bphi_i, 	\\
			&  \text{and~} 0  = \mC(\Bs_i,\bsigma_i) = \mC_i .
		\end{cases} 
	\label{eq:optimalmodif_discrete}
	\end{equation} 
	where we use the short-hand notation $ L_i = L_{lof}\left( \Bs_i , \bsigma_i \right)$. $\phi$ and $\mC$ are prescribed functions of their arguments. 
	If $\Bs$ is the inverse temperature vector, the state equation reads $\phi_l(\Bs_i,\bsigma_i) = s_{i,l} \sigma_{i,l}$ and  $\mC(\Bs_i,\bsigma_i) = -\bsigma_i \cdot \Bs_i^{-1}$.
	In practice, a convenient choice is to take the state as the neg-entropy vector, namely $\Bs = \log \bb/\beta$. This choice guarantees that the algorithm will not optimize over non-realizable negative inverse temperature vectors.	
	if $\Bs$ is the neg-entropy, the state equation is $\phi_l(\Bs_i,\bsigma_i) = \sigma_{i,l}$ and the energy dissipation constraint reads  $\mC(\Bs_i,\bsigma_i) = \bsigma_i \cdot e^{-\Bs_i}$.
	
	\paragraph{The Discrete augmented cost and its gradient.}
	To solve the optimization problem (\ref{eq:optimalmodif_discrete}), we define the discrete augmented cost as a counterpart to (\ref{eq:extentedlag_withweight})~: 
		\begin{equation}
		\begin{split}
			& \mJ_{\lambda,\mu}[\bsigma] = \sum_{i=0}^{\nt-1} \dfrac{\bpi_i + \bpi_{i+1}}{2} \cdot \left(\Bs_{i+1} - \Bs_i\right) -h \dfrac{H_i+H_{i+1}}{2},\\
			&\text{where~} H_i =   \bpi_i \cdot \bphi_i - L_i + \lambda_i \mC_i - \dfrac{1}{2\mu}\mC_i^2 = \mH_{P,\lambda_i,\mu} (\Bs_i,\bpi_i,\bsigma_i).	
		\end{split}
		\label{eq:extentedlag_withweight_discrete}
		\end{equation}
denotes the discrete counterpart to the augmented Hamilton-Pontryagin function.
For the notation (\ref{eq:extentedlag_withweight_discrete}) to be self-consistent, and for the cost to be a function of the control only, the state and the co-state need to be enslaved to it. The good prescription is to define the states and the co-states recursively as follows~:
\begin{equation}
	\begin{split}
	 \Bs_0 = \Bs[t=0], ~  \Bs_{1} =  \Bs_{0} + \Dt \nabla_{\bpi_0} H_0 \text{~~and~~} \Bs_i = \Bs_{i-1} + 2 \Dt \nabla_{\bpi_i} H_i \text{~~($i>1$)};\\
	  \pi_\nt = 0, ~  \bpi_{\nt-1} =  \Dt \nabla_{\Bs_\nt} H_\nt \text{~~and~~} \bpi_{i-1} = \bpi_{i+1} + 2 \Dt \nabla_{\Bs_i} H_i \text{~~($i<\nt-1$)}.
	\end{split}
	\label{eq:optimalestimates_discrete}
\end{equation}
Those two equations are the discrete counterpart to Equation (\ref{eq:optimalestimates}).
The gradient of the augmented cost with respect to the $\nt \times K $ control variables $\sigma_{i,l}$ is then obtained as 
\begin{equation}
\nabla_\bsigma \mJ_{\lambda,\mu} = \left(\dfrac{\partial \mJ_{\lambda,\mu}}{\partial \sigma_{i,l}} \right)_{\substack{0\le i \le \nt \\ 1 \le l \le K}}
\text{with } \dfrac{\partial \mJ_{\lambda,\mu}}{\partial \sigma_{i,l}} =
\begin{cases}
& -\Dt \dfrac{\partial \mH_i}{\partial \sigma_{i,l}} \text{~~if $0<i<\nt$}, \\
&  -\dfrac{\Dt}{2} \dfrac{\partial \mH_i}{\partial \sigma_{i,l}} \text{~~if $i=0$ or $i = \nt$}.\\
\end{cases}
\label{eq:discretegradient}
\end{equation}

To see that the prescription (\ref{eq:optimalestimates_discrete}) is correct and leads to the gradient (\ref{eq:discretegradient}), it suffices  to obtain
the infinitesimal variation of the cost with respect to the state, co-state and control variables as : 
\begin{equation*}
\begin{split}
	2 \delta \mJ  = & \sum_{i=1}^{\nt-1}  \delta \bpi_i \cdot \left( \Bs_{i+1} - \Bs_{i-1} -2 \Dt \nabla_{\bpi_i} H_i\right) + \delta \Bs_i \cdot \left( \bpi_{i-1} - \bpi_{i+1} -2 \Dt \nabla_{\Bs_i} H_i\right) \\
	& + \delta \bpi_\nt \cdot \left( \Bs_{\nt} - \Bs_{\nt-1} - \Dt \nabla_{\bpi_\nt} H_\nt\right) +  \delta \bpi_0 \cdot \left( \Bs_{1} - \Bs_{0} - \Dt \nabla_{\bpi_0} H_0\right) \\
	& + \delta \Bs_\nt \cdot \left( \bpi_{\nt} + \bpi_{\nt-1} - \Dt \nabla_{\Bs_\nt} H_\nt\right) +  \delta \Bs_0 \cdot \left( -\bpi_0  - \bpi_{1} - \Dt \nabla_{\Bs_0} H_0\right) \\
	& - 2 \sum_{i=1}^{\nt-1}  \Dt \delta \bsigma_i  \cdot \nabla_{\bsigma_i} H_i- \Dt \delta \bsigma_0 \cdot \nabla_{\bsigma_0} H_0  - \Dt \delta \bsigma_\nt \cdot \nabla_{\bsigma_\nt } H_\nt.
\end{split}
\end{equation*}
With the prescription (\ref{eq:optimalestimates_discrete}),  the first three lines of the previous expression vanish. The last line leads directly to (\ref{eq:discretegradient}).

\paragraph{Norms.}
To control the convergence of the numerics and define stopping criteria, we define norms for both the cost gradient (\ref{eq:discretegradient}) and the constraint. 
We use for the gradient~:
\begin{equation}
|| \nabla_\sigma \mJ || = \dfrac{1}{K} \left(\sum_{i=0}^{\nt} \sum_{l=1}^K \left( \dfrac{\partial \mJ_{\lambda,\mu}}{\partial \sigma_{i,l}} \right)^2\right)^{1/2}
\label{eq:discretegradientnorm}
\end{equation}
As for the constraint satisfiability, that determines how well the energy is conserved~: 
\begin{equation}
|| \mC || = \sup_{0\le i \le \nt} |\mC_i|.
\label{eq:discretegradientnorm}
\end{equation}

\subsection{Implementation of the optimal response algorithm.}
\paragraph{Augmented Lagrangian descent.}
\newcommand{\aone}{{\alpha_1}}
\newcommand{\atwo}{{\alpha_2}}
\newcommand{\gtol}{g_{tol}}
\newcommand{\ctol}{c_{tol}}
\newcommand{\gttol}{\tilde g_{tol}}
\newcommand{\cttol}{\tilde c_{tol}}

 Algorithm \ref{algo:lancelot} describes  the  augmented Lagrangian descent method involved in the  optimal response algorithm (\ref{sec:Pontryalgo}). As previously explained, the augmented Lagrangian method provides a strategy to determine iteratively the Lagrange multiplier $\lambda$ and the scalar penalty factor $\mu$, so that the norms of both the constraint and the cost gradient utlimately become lower than some prescribed convergence levels  $\ctol$ and $\gtol$. We use a simplified version of the LANCELOT method of multipliers described in \cite{nocedal_numerical_2006} (Chapter 17, algorithm 17.4). In a nutshell, it consists in decreasing $\mu$ when the constraint is not sufficiently satisfied and updating $\lambda $ otherwise, so that the latter mimics the behavior of Lagrange multiplier.
\begin{algorithm}[h]
\caption{Augmented descent with the LANCELOT method of multipliers}\label{algo:lancelot}
\begin{algorithmic}[1]
\STATE $\mu \leftarrow \mu^{(0)}$ \hfill  \COMMENT{Initialize the weigths}
\STATE $ \lambda \leftarrow \lambda^{(0)} $ 
\STATE $\gttol \leftarrow \max \,\left \lbrace \mu,\,\gtol \right \rbrace$  \hfill  \COMMENT{Initialize loosened tolerances}
\STATE  $\cttol \leftarrow \max \, \left \lbrace \mu^{\atwo},\,\ctol \right \rbrace$ 
	\WHILE {$||\nabla_\bsigma J|| > \gtol $ or $||\mC|| > \ctol$}
		\STATE  Determine $\bsigma_{\lambda,\mu}^\star = \arg \inf  J_{\lambda,\mu}$ with precision $\gttol$ \hfill  \COMMENT{Unconstrained Descent}
		\IF{$||\mC || < \ctol $ }
			\STATE $ \lambda_i \leftarrow \lambda_i - \dfrac{\mC_i}{\mu}$ 	\hfill  \COMMENT{Update $\lambda$}
			\STATE $\cttol \leftarrow \max \left \lbrace \cttol \mu^\aone\,, \,\ctol  \right \rbrace$	 \hfill  \COMMENT{Tighten tolerances}
			\STATE $\gttol \leftarrow \max \left \lbrace \gttol \mu\,, \,\gtol \right \rbrace$
		\ELSE
			\STATE $\mu \leftarrow \tau \mu $ with $\tau = \min \left \lbrace 0.2, \, \sqrt \mu \right \rbrace $ \hfill  \COMMENT{Update $\mu$}
			\STATE $\cttol \leftarrow \max \left \lbrace \cttol \mu^\atwo, \,\ctol \right \rbrace$ \hfill \COMMENT{Tighten tolerances}
			\STATE $\gttol \leftarrow \max \left\lbrace \mu,\gtol \right \rbrace$
		\ENDIF
	\ENDWHILE
\RETURN $\bsigma_{\lambda,\mu}^\star$ \hfill \COMMENT{Return the optimal control}
\end{algorithmic}
\end{algorithm}
For all of our numerics, we used the empirical values  $\aone=0.9$ and $\atwo= 0.1$ to control the tightening of the tolerances.
If no initial guess was provided, we used $\mu^{(0)} = 0.1$ and  $\lambda^{(0)} = 0$.

\paragraph{Unconstrained descent.}
The success of the augmented descent method is dependent upon having an efficient unconstrained descent solver, to determine the minimum of the augmented cost $J_{\lambda,\mu}$ up to the desired tolerances $\gttol$ and $\cttol$. We found that  the non-linear conjugate gradient method with Polak-Ribi\`ere$^+$ updates (see Chapter 5 and Formula (5.43) of \cite{nocedal_numerical_2006}) is here suited for the task. This is a standard and well-documented method, which we describe in Algorithm (\ref{algo:nlcg}) for the sake of clarity.
\footnote{In Algorithm (\ref{algo:nlcg}), we abuse the use of our previously defined notation $\cdot$. The descent direction and the cost gradient are vectors that take values in $\mathbb R^{K} \times \mathbb R^{\nt+1}$, and the notation $\ba \cdot \bp $ there stands for  $\sum_{i=0}^\nt \sum_{l=1}^K a_{i,l} p_{i,l}$.}

  \begin{algorithm}[h]
  \newcommand{\cost}{\mJ_{\lambda,\mu}}
\caption{The non-linear conjugate gradient descent}\label{algo:nlcg}
\begin{algorithmic}[1]
\STATE $\bsigma  \leftarrow  \bsigma_{\lambda,\mu}^{(0)} $ \hfill \COMMENT{Initial guess}
\STATE  $J  \leftarrow \cost(\bsigma,t)$ 		\hfill \COMMENT{From  Equations (\ref{eq:optimalestimates_discrete}) and (\ref{eq:extentedlag_withweight_discrete})}
\STATE $\nabla J  \leftarrow   \nabla_\bsigma\cost(\bsigma)$ \hfill \COMMENT{Using Equation (\ref{eq:discretegradient})}
\STATE $\bp  \leftarrow  -\nabla J$ \hfill \COMMENT{Initial guess for the descent direction}
\WHILE{$||\nabla J|| > \gtol$}
	\STATE  Find $\alpha$ such  that  $\bsigma + \alpha \bp$  satisfies the strong Wolfe Conditions, namely
	\begin{equation}
	J(\bsigma + \alpha \bp) \le J + c_1 \alpha \nabla J \cdot \bp \text{~and~ }\nabla_\bsigma\cost(\bsigma + \alpha \bp ) \cdot \bp \ge c_2 \nabla J \cdot \bp.
	\label{eq:strongwolfe}
	\end{equation}
	\STATE $\bsigma   \leftarrow \bsigma+ \alpha \bp $.
	\STATE  $J  \leftarrow \cost(\bsigma,t)$ 
	\STATE $\nabla J^\prime   \leftarrow \nabla J$  and $\nabla J  \leftarrow   \nabla_\bsigma\cost(\bsigma,t)$ 
	\STATE $\bp  \leftarrow \gamma \bp-\nabla J $ with $\gamma = \max \left\lbrace\gamma^{PR},0 \right \rbrace$ and $\gamma^{PR}  = \dfrac{\nabla J \cdot \left ( \nabla  J - \nabla J^\prime \right)}{\nabla J^\prime  \cdot \nabla J^\prime}$
\ENDWHILE
\RETURN $\bsigma$
\end{algorithmic}
\end{algorithm}
The non-linear conjugate gradient descent uses yet  another layer of optimization. As apparent on on lines 6 and 7 of Algorithm (\ref{algo:nlcg}), it requires to perform  a unidimensional optimization, in order to  find a  scalar $\alpha$ that verifies the so-called  strong Wolfe conditions (\ref{eq:strongwolfe}), for prescribed parameters $c_1$ and $c_2$.   Fortunately, this kind of optimization is very standard, and a suited value for the scalar $\alpha$ can be found using the default line-search algorithms implemented in most programming languages. In this work, we used the  Python programming language (Python Software Foundation, https://www.python.org/). The line-search function available from the scipy.optimize package implements Algorithm 3.2 of \cite[Chapter 3 ]{nocedal_numerical_2006}) and performs a unidimensional optimization thats finds a  scalar $\alpha$ that verifies the strong Wolfe conditions. In our numerics, we have used $c_1=10^{-3}$ and $c_2=0.5$.

 \section*{Acknowledgments}
The work reported in this paper was partially supported by the National Science Foundation under grant DMS-1312576.

\section*{References}
\bibliographystyle{unsrt}
\bibliography{FDBiblio}
\end{document}